\documentclass[conference,compsoc]{IEEEtran}
\ifCLASSOPTIONcompsoc
  \usepackage[nocompress]{cite}
\else
  \usepackage{cite}
\fi
\usepackage{tikz}
\usepackage{amsmath}

\usepackage{hyperref} 

\makeatletter
\g@addto@macro{\UrlBreaks}{\do\-}
\makeatother

\usepackage{cite}
\usepackage{amsmath,amssymb,amsfonts}
\usepackage{algorithm}
\usepackage{algpseudocode}
\usepackage{graphicx}
\usepackage{textcomp}
\usepackage{xcolor}
\usepackage{xspace}
\usepackage{kotex}
\usepackage{lipsum}
\usepackage{enumitem}
\usepackage{fancyhdr}
\usepackage{hyperref}
\usepackage{comment}
\usepackage{makecell}
\usepackage{tabularx}
\usepackage{multirow}
\usepackage{comment}
\usepackage{xfrac}
\usepackage{multirow}
\usepackage{tcolorbox}
\usepackage{booktabs}
\usepackage{color,soul}
\usepackage[utf8]{inputenc}
\usepackage[english]{babel}
\usepackage{balance}
\usepackage{circledsteps}
\usepackage{tikz}
\usepackage{chngcntr} % for appendix figure numbering
\usepackage{wasysym} % for circles
\usepackage{textcase} % for lowercase forcing

\usepackage{tikz}

\newcommand\copyrighttext{%
  \footnotesize \textcopyright \the\year{} IEEE. Personal use of this material is permitted. Permission from IEEE must be obtained for all other uses, including reprinting/republishing this material for advertising or promotional purposes, collecting new collected works for resale or redistribution to servers or lists, or reuse of any copyrighted component of this work in other works.}

\newcommand\copyrightnotice{%
\begin{tikzpicture}[remember picture,overlay]
\node[anchor=south,yshift=10pt] at (current page.south) {\fbox{\parbox{\dimexpr0.75\textwidth-\fboxsep-\fboxrule\relax}{\copyrighttext}}};
\end{tikzpicture}%
}

\newcommand{\reservoir}{Reservoir-Sampling\xspace}
\newcommand{\stickysampling}{Sticky-Sampling\xspace}
\newcommand{\misragries}{Misra-Gries\xspace}
\newcommand{\spacesaving}{Space-Saving\xspace}
\newcommand{\lossycounting}{Lossy-Counting\xspace}
\newcommand{\cmsketch}{CountMin-Sketch\xspace}
\newcommand{\csketch}{Count-Sketch\xspace}

\newcommand{\minentry}{\texttt{Min}\xspace}

\newcommand{\psample}{$P_{sample}$\xspace}

\newcommand{\averagecase}{\texttt{average-case}\xspace}
\newcommand{\worstcase}{\texttt{worst-case}\xspace}

\newcommand{\update}{\texttt{Update}\xspace}
\newcommand{\compress}{\texttt{Compress}\xspace}

\newcommand{\rhth}{$\text{RH}_\text{TH}$\xspace}
\newcommand{\rhtheq}{\text{RH}_\text{TH}\xspace}
\newcommand{\fest}{$f_{est}$\xspace}
\newcommand{\festeq}{f_{est}\xspace}
\newcommand{\freal}{$f_{real}$\xspace}
\newcommand{\frealeq}{f_{real}\xspace}

\newcommand{\us}{$\mu\text{s}$\xspace}
\newcommand{\umsquare}{$\mu\text{m}^\text{2}$\xspace}

\newcommand{\eg}{\emph{e.g.}\xspace}
\newcommand{\ie}{\emph{i.e.}\xspace}

\newcommand{\blackcircled}[1]{%
    \tikz[baseline=(char.base)]{
        \node[shape=circle, draw, fill=black, inner sep=1pt] (char) {\textcolor{white}{\small #1}};
    }%
}

\newcommand*{\Scale}[2][4]{\scalebox{#1}{$#2$}}
\begin{document}
\title{SoK: Systematizing a Decade of Architectural RowHammer Defenses\\Through the Lens of Streaming Algorithms}

\author{
\IEEEauthorblockN{
Michael Jaemin Kim\IEEEauthorrefmark{4},
Seungmin Baek\IEEEauthorrefmark{2},
Jumin Kim\IEEEauthorrefmark{2},
Hwayong Nam\IEEEauthorrefmark{2},
Nam Sung Kim\IEEEauthorrefmark{3},
Jung Ho Ahn\IEEEauthorrefmark{2}}
\IEEEauthorblockA{\IEEEauthorrefmark{4}Meta, \IEEEauthorrefmark{2}Seoul National University, \IEEEauthorrefmark{3}University of Illinois at Urbana-Champaign}
\IEEEauthorblockN{
\IEEEauthorrefmark{4}michael604@meta.com,
\IEEEauthorrefmark{2}\{qortmdalss, tkfkaskan1, nhy4916, gajh\}@snu.ac.kr, 
\IEEEauthorrefmark{3}nskim@illinois.edu
}
}


\maketitle

\copyrightnotice

\begin{abstract}
A decade after its academic introduction, RowHammer (RH) remains a moving target that continues to challenge both the industry and academia.
With its potential to serve as a critical attack vector, the ever-decreasing RH threshold now threatens DRAM process technology scaling, with a superlinearly increasing cost of RH protection solutions.
Due to their generality and relatively lower performance costs, architectural RH solutions are the first line of defense against RH.
However, the field is fragmented with varying views of the problem, terminologies, and even threat models.

In this paper, we systematize architectural RH defenses from the last decade through the lens of streaming algorithms.
We provide a taxonomy that encompasses 48 different works.
We map multiple architectural RH defenses to the classical streaming algorithms, which extends to multiple proposals that did not identify this link.
We also provide two practitioner guides.
The first guide analyzes which algorithm best fits a given \rhth, location, process technology, storage type, and mitigative action.
The second guide encourages future research to consult existing algorithms when architecting RH defenses.
We illustrate this by demonstrating how \reservoir can improve related RH defenses, and also introduce \stickysampling that can provide mathematical security that related studies do not guarantee.

\end{abstract}


% \fi
\IEEEpeerreviewmaketitle

\section{Introduction}
\label{sec:1_introduction}

\begin{figure}[!tb]
  \center
  \vspace{0.0in}
  \includegraphics[width=0.9\linewidth]{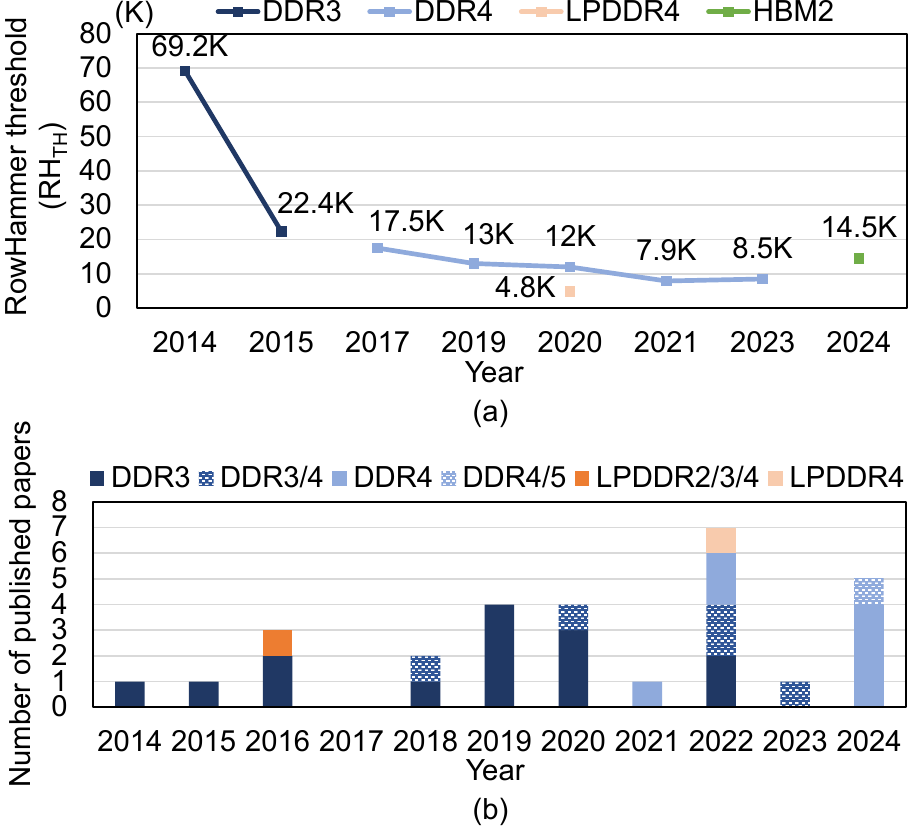}
  \vspace{-0.05in}
  \caption{
  (a) The trend of RowHammer threshold~\cite{arxiv-2024-combined, isca-2020-revisiting, micro-2021-uncovering, dsn-2024-hbmrowhammer},
  (b) the number of RowHammer attack publications from S\&P, USENIX Security, CCS, USENIX ATC, Black Hat, MICRO, and ISCA.
  }
  \vspace{-0.07in}
  \label{fig:research_trend}
\end{figure}

As DRAM process technology scales, the interference between nearby DRAM cells intensifies.
One of its consequences is RowHammer (RH)~\cite{isca-2014-flipping}; when a DRAM row (aggressor) is frequently activated, surpassing a certain threshold (\rhth), the DRAM cells in the nearby row (victim) gradually experience bitflips.
Since its academic disclosure in 2014~\cite{isca-2014-flipping}, RH has served as a powerful primitive for various attacks on diverse environments, from privilege escalation in web browsers~\cite{dimva-2016-rowhammerjs, sec-2021-smash} to the machine learning model inference depletion~\cite{security-2024-tossing,security-2024-yes,sp-2022-deepsteal,sec-2020-deephammer,sec-2019-terminalbrain}, and many more~\cite{atc-2018-throwhammer,blackhat-2015-privilege,ccs-2016-drammer,sec-2022-halfdouble,sp-2018-anotherflip,micro-2020-pthammer,sec-2016-onebitonecloud,sosp-2017-sgxbomb,sec-2024-gadgethammer,sp-2020-rambleed,sec-2020-deephammer,sp-2022-deepsteal,sp-2022-specHammer,sec-2016-fengshui,asplos-2025-marionette} (Figure~\ref{fig:research_trend}).
Further, RH is becoming a reliability problem as \rhth decreases~\cite{isca-2020-revisiting}, where benign workloads cause RH bitflips in certain scenarios~\cite{isca-2022-moesiprime}.

As a response, numerous architectural RH defense solutions (hereafter, architectural solutions) have been proposed.
We found 48 different proposals from various venues such as S\&P, USENIX Security, ISCA, MICRO, ASPLOS, HPCA, IEEE CAL, DAC, DATE, and ICCD.
They were proposed across over the past decade, during which \rhth has decreased by a factor of over 40 and new characteristics such as RowPress~\cite{isca-2023-rowpress} was discovered.
Naturally, this field became fragmented; each solution presents novel yet diverse views of the problem, with its own terminologies, target security level, threat model, making it difficult to compare their area, energy, and performance overheads.

We pursue providing a \textbf{systematization of the architectural RH defenses; especially with the lens of \emph{streaming algorithms}}.
Streaming algorithms~\cite{FTTCS-2005-muthukrishnan-textbook} handle a stream of data, analogous to a stream of activation commands in the RH context.
They are a well-established and rich field, already widely employed in the network~\cite{SIGCOMM-2002-Sticky-sampling-weaker} or database~\cite{ICDT-2005-Space-Saving} studies.
For example, an algorithm we discuss later (\S\ref{subsec:reservoir_sampling}), \reservoir, dates back to 1962~\cite{JASA-1962-Method-4}, with a wide range of optimizations~\cite{TOMS-1985-algorithm-Z,TOMS-1994-algorithm-L}.
The domain-level connection between streaming algorithms and RH defense was previously identified~\cite{micro-2020-graphene}.
We extend this understanding, mapping the majority of the existing RH protection to the relevant algorithms.
Our contributions are as follows:

\noindent
\textbf{1) Systematizing architectural RH defense solutions (\S\ref{sec:3_taxonomy}).}
We first categorize the architectural solutions into three types; tracker-based, tracker-less, and cooperative.
For both defenses of tracker-based and tracker-less, we systematize each solution regarding their main goal, related streaming algorithm, mitigative action (refresh/throttle/shuffle), security guarantee (deterministic/probabilistic/empirical), which process technology they employed (logic/memory), and architectural optimizations.
Also, we delve into the general goal of cooperative defense solutions, which works together with either tracker-based or tracker-less defenses.
Lastly, we discuss how architectural techniques have been synergistically used with algorithmic trackers for various defenses.
We identify multiple new connections of certain prior RH defenses to various streaming algorithms;
CBT~\cite{cal-2017-CBT,isca-2018-CBT} for Counter-trie~\cite{imc-2004-Counter-trie}, TWiCe~\cite{isca-2019-twice} for \lossycounting~\cite{VLDB-2002-Lossy-sticky}, DBC~\cite{CAL-2019-DBC} for the sliding window~\cite{PODS-2004-approximate-counts-Arvind}, DAPPER~\cite{hpca-2025-dapper} for Yogev et al.~\cite{JACM-2022-framework}, and PARFM~\cite{hpca-2022-mithril}/PrIDE~\cite{isca-2024-pride}/MINT~\cite{MICRO-2024-mint} for \reservoir~\cite{JASA-1962-Method-4}.

\noindent
\textbf{2) Which algorithm to use when (\S\ref{sec:4_good_combination})?}
We provide guides in choosing the most suitable algorithm for given circumstances.
For tracker-based defense, we particularly investigate the area overhead with regard to the type of a streaming algorithm, target \rhth, the process technology, and the type of tracker (\eg, CAM/SRAM).
We focus on the area overhead, because it has been the first-class citizen in tracker-based defenses.
For example, the proprietary RH defenses employed on commodity devices aggressively minimized resource usage~\cite{micro-2021-uncovering,arxiv-2023-dsac,isscc-2023-hynixrowhammer}, becoming vulnerable to sophisticated attacks~\cite{sp-2020-trrespass,sp-2022-blacksmith}.

We explain suitable pairing between the type of mitigative action and streaming algorithm, based on the type of error bounds the algorithm provides.
Finally, we summarize several architectural optimization approaches that can be synergistically used with algorithmic trackers in the future: caching, anomaly detection, spatial variation of RH vulnerability, sliding window considering periodic refresh, and new memory protocols (e.g., Compute Express Link (CXL)).

\noindent
\textbf{3) Case Studies of Prior Defenses and Algorithms (\S\ref{sec:check_existing_algorithm}).}
Finally, we provide case studies of several prior RH defenses and their corresponding algorithms, revealing previously unrecognized connections.
As our newly found connections with streaming algorithms suggest, it is not uncommon for RH defenses to work on a similar problem that existing algorithms solve.\footnote{Such cases occur even within the domain of streaming algorithm. \misragries~\cite{scp-1982-misragries} has been reinvented multiple times~\cite{TODS-2003-Misra-Gries-reinvent1,ESA-2002-misra-gries-reinvent2}, sometimes in its isomorphic form~\cite{ICDT-2005-Space-Saving,agarwal-2013-mergeable}.}
As an example, we demonstrate that \reservoir~\cite{JASA-1962-Method-4} can provide more \emph{efficient} defense solution over the related PARFM/PrIDE/MINT, for the first time.
Further, we illustrate how tracker-based defenses that are augmented with sampling~\cite{dac-2017-prohit,dac-2019-mrloc,arXiv-2024-proteas,arxiv-2023-dsac} have historically provided only empirical security guarantees, which were often broken over time.
As an example, we demonstrate two phase-changing adversarial patterns against each technique of DSAC~\cite{arxiv-2023-dsac} for the first time.
The security of the DSAC with its combined technique is ambiguous, which proved to be vulnerable to a fuzzer~\cite{isca-2024-pride}.
Regarding such an approach of merging sampling with trackers, we introduce \stickysampling~\cite{VLDB-2002-Lossy-sticky} algorithm, for the first time in RH context.
Unlike other related solutions~\cite{dac-2017-prohit,dac-2019-mrloc,arxiv-2023-dsac,arXiv-2024-proteas}, the algorithm is equipped with a compression phase and provides a mathematical bound for security guarantee.

\section{Background}
\label{sec:2_background}

\subsection{Related Work}
\label{sec:7_related_work}

We do not systematize software-based RH defenses in this paper.
Despite their effectiveness and deployability, they either lack scalability when \rhth drops or have limited scope.
Software tracking-based defenses~\cite{asplos-2016-anvil,asplos-2024-tarot,atc-2022-softtrr} prevent RH bitflips by sending additional refreshes to the victim rows, while mitigating performance costs through system optimization.
Software isolation-based defenses~\cite{osdi-2018-zebram,security-2017-catt,dimva-2018-guardion,sosp-2023-siloz, asplos-2019-cta} leverage guard rows to create a physical gap and protect kernel data or prevent inter-VM (virtual machine) RH attacks.
Such software-based defenses can be orthogonally applied with architectural RH defenses.

Zhang et al.~\cite{asiaccs-rowhammer-sok} systematize the RH attacks and defenses.
However, the work is focused on the RH attacks and software-based defenses, with brief discussion of the ECC-based defenses.
Kim et al.~\cite{isca-2020-revisiting} provide valuable insights on the trend of \rhth and their effects on the architectural RH defenses.
However, the study does not cover recent advances in RH defenses or the investigation of the related streaming algorithms.
We comprehensively systematize 48 architectural RH defenses, with an effort to better understand the field through the lens of streaming algorithms.

%-------------------------------------------------------------------------------
\subsection{Basic DRAM Operation and RH}
%-------------------------------------------------------------------------------
\label{subsec:2_dram_operation}

DRAM activates a row of cells with the ACT command (\texttt{ACT}) before accessing them.
The subsequent \texttt{ACT}s on the same DRAM bank require a tRC (\eg, 48ns) gap between them.
Due to the inherent charge loss of a DRAM cell over time, a host memory controller (MC) sends a refresh command (\texttt{REF}) to the DRAM device.
\texttt{REF} is sent periodically, typically at every tREFI (\eg, 3.9\us), each with tRFC (\eg, 350ns) time window to execute refresh on cells.
JEDEC specification defines that each DRAM row is refreshed at least once every tREFW (\eg, 32ms).
Terminologies are summarized in Table~\ref{tbl:symbols_background}.

Repeatedly activating a DRAM row (aggressor) causes electromagnetic disturbance on the nearby rows (victim).
Typically, the victim row can experience a bitflip when the number of activations exceeds a certain value of \rhth.
There also exists another form of vulnerability, RowPress~\cite{isca-2023-rowpress}, where opening the row for a long time results in a similar effect.
We only consider \rhth unless stated otherwise, for the sake of clarity.
The RowPress effect can also be incorporated into the \rhth model~\cite{micro-2024-impress}.
Increasing activations can also increase the number of bit-flips.
Because such errors can quickly occur in a bursty way, even a strong form of ECC may not be able to correct all RH-induced errors~\cite{sp-2019-eccploit, micro-2023-cube, ndss-2023-cof}.

DRAM manufacturers have adopted proprietary RH defenses that utilize a certain time margin from existing \texttt{REF}s~\cite{micro-2021-uncovering,arxiv-2023-dsac}.
Recent JEDEC specification also defines RFM (refresh management) and PRAC (per-row activation counting) commands, which grant an additional timing margin to the DRAM device based on the rate of \texttt{ACT}s~\cite{jedec-2024-ddr5} or when the DRAM-side RH solution desires~\cite{jedec-2024-ddr5}.

\begin{table}[!tb]
    \centering
    \small
    \caption{Symbols and terminologies in \S\ref{sec:2_background}}
    \label{tbl:symbols_background}
    \Scale[0.95]{
    \begin{tabular}{p{0.27\columnwidth} p{0.67\columnwidth}} 
        \Xhline{3\arrayrulewidth}
        \textbf{Symbol} & \textbf{Description} \\ 
        \Xhline{1.5\arrayrulewidth}
        \rhth   & RowHammer threshold. \\
        tREFI   & Time between two consecutive REF commands (\eg, 3.9\us). \\
        tREFW   & Time window where each row is guaranteed to be refreshed at least once (\eg, 32ms). \\ 
        \worstcase      & Streaming algorithms provide complexities and bounds that hold for the worst case. \\
        \averagecase    & The efficacy of an algorithm may differ on the average case real-world data. \\
        %asdf    & asdf \\
        %asdf    & asdf \\
        \Xhline{0.5\arrayrulewidth}
        \Xhline{3\arrayrulewidth}
    \end{tabular}
    }
    \vspace{-0.08in}
\end{table}

%-------------------------------------------------------------------------------
\subsection{RH Attacks and Solutions}
%-------------------------------------------------------------------------------
\label{subsec:2_rh_attack}

Since its academic disclosure in 2014 on DDR3~\cite{isca-2014-flipping}, RH served as a hard-to-prevent critical attack vector, enabling various sophisticated attacks~\cite{dimva-2016-rowhammerjs, sec-2021-smash,security-2024-tossing,security-2024-yes,sp-2022-deepsteal,sec-2020-deephammer,sec-2019-terminalbrain,atc-2018-throwhammer,blackhat-2015-privilege,ccs-2016-drammer,sec-2022-halfdouble,sp-2018-anotherflip,micro-2020-pthammer,sec-2016-onebitonecloud,sosp-2017-sgxbomb,sec-2024-gadgethammer,sp-2020-rambleed,sec-2020-deephammer,sp-2022-deepsteal,sp-2022-specHammer,sec-2016-fengshui,asplos-2025-marionette}.
To tackle this, various architectural solutions have been proposed.
At the same time, DRAM manufacturers adopted proprietary RH solutions on DDR4 and other DRAM devices (\eg, in-DRAM TRR~\cite{micro-2021-uncovering,arxiv-2023-dsac,isscc-2023-hynixrowhammer}).
However, despite their claims at the time, such solutions have proven to be insecure against adversarial access patterns~\cite{sp-2020-trrespass,sp-2022-blacksmith, micro-2021-uncovering}.

Despite a decade of research, the number of research papers on RH attacks is not declining and RH remains a serious threat (Figure~\ref{fig:research_trend}(b)).
In-DRAM TRR or ECCs provide only limited security or robustness against RH attacks~\cite{sp-2019-eccploit}.
Recent studies discover techniques where even a few RH bitflips can compromise the reliability and security of whole machine learning models~\cite{security-2024-tossing,security-2024-yes, sp-2022-deepsteal,sec-2020-deephammer,sec-2019-terminalbrain}.

\rhth has also continuously decreased with the process technology scaling (Figure~\ref{fig:research_trend}(a)).
Even benign workloads experiencing a large number of activations can cause RH bitflips~\cite{isca-2022-moesiprime} under special circumstances.
Considering that the area, energy, and performance overheads of RH solutions often superlinearly increase as \rhth decreases, it is a serious burden on the DRAM manufacturers.
Without continuously optimizing the architectural defenses to minimize such costs, they can again turn to less-secure proprietary RH solutions and rely on the security through obscurity.
%

%-------------------------------------------------------------------------------
\subsection{Streaming Algorithms}
%-------------------------------------------------------------------------------
\label{subsec:2_algorithms}

Streaming algorithms~\cite{FTTCS-2005-muthukrishnan-textbook} 
handle a stream of data to extract useful information, with limited space and compute resources.
Such algorithms typically operate in a single pass, processing each stream element once without revisiting past elements.
They have been studied in depth and widely used in the context of networking~\cite{SIGCOMM-2002-Sticky-sampling-weaker
} and databases~\cite{ICDT-2005-Space-Saving}.
For example, the ``heavy-hitter'' algorithm~\cite{FTTCS-2005-muthukrishnan-textbook} can identify network IP addresses that generate a heavy load of requests, to allocate appropriate resources~\cite{Infocom-2006-Sketch-guided-sampling}, or to prevent DoS attacks~\cite{SIGCOMM-2002-Sticky-sampling-weaker}.

\begin{figure}[!tb]
  \center
  \vspace{0.0in}
  \includegraphics[width=0.98\linewidth]{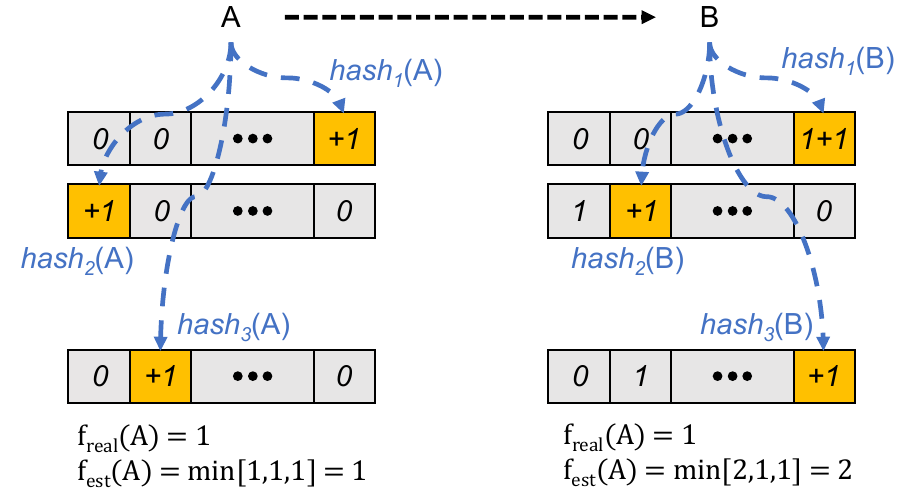}
  \vspace{-0.05in}
  \caption{Example diagram of \cmsketch when processing element A, then B. $f_{est}(A)$ is over-estimated due to the hash collision of $hash_1(B)$.}
  \vspace{-0.07in}
  \label{fig:CMS}
\end{figure}

\noindent
\textbf{Frequency estimation algorithm.}
Although the goals of various streaming algorithms are diverse~\cite{FTTCS-2005-muthukrishnan-textbook} (\eg, quantile, frequent moment, or cardinality), one of the most studied tasks is \emph{heavy-hitter} identification, or more generally, \emph{frequency estimation}.
The frequency estimation algorithm approximates the number of occurrences of an item in the observed stream elements, with a certain mathematical bound with regard to the real frequency.

\vspace{-0.05in}
\Scale[1.0]{
\begin{minipage}{1.0\columnwidth}
\begin{align*}
& \festeq - error \overset{\delta}{\leq} \frealeq \overset{\delta}\leq \festeq + error
\end{align*}
\end{minipage}}\\
\vspace{0.05in}

As noted above, we denote the estimated frequency as \fest, and real frequency as \freal.
The precise definition of the ``\emph{error}'' term varies by algorithm. 
$\delta$ is a confidence factor of the inequation.
When an algorithm is deterministic, both left and right $\delta$ is 1.
For a probabilistic algorithm, the $\delta$ can be a value between $0$ to $1$, which holds for certain assumptions on the randomization and the stream pattern.

\noindent\textbf{\cmsketch example.}
We provide an example of \cmsketch to elaborate the frequency estimation algorithm (Figure~\ref{fig:CMS}).
An experienced reader can skip this.
\cmsketch holds multiple counter rows with a unique hash function per row.
Each element in a stream is processed by \cmsketch one-by-one.
The address of the element is mapped to one counter per row, using the unique hash function per row.
At the end of the stream, \cmsketch estimates the frequency of a certain element as the \emph{minimum} among its mapped counters from each row.
In this case, \fest can be expressed as below.

\vspace{-0.15in}
\begin{center}
\begin{minipage}{1.0\columnwidth}
\begin{align*}
  & \text{\cmsketch:}\ \ \frealeq \leq \festeq 
    \overset{1-\delta}\leq \frealeq + \epsilon\|f_{-a}\|_1
\end{align*}
\end{minipage}
\end{center}
\vspace{0.05in}

The inequation on the left hand side is deterministic ($\delta =1$) as once an element increments a counter it is not deleted.
The right-hand-side one is probabilistic because hash collision can bloat the estimated frequency.
Such a probability is based on the assumptions that the stream is oblivious and the hash function is random.
An attacker with knowledge of the hash function (an adaptive attacker) can craft an access pattern that significantly skews this right-hand side inequation with a crafted access pattern~\cite{PODS-2020-adversarial}.

\noindent
\textbf{Sampling algorithm.}
While sampling can be a subroutine for a larger streaming algorithm for frequency estimation (\eg, \stickysampling), it can also be considered as a streaming algorithm.
For example, sampling is a part of streaming algorithms, such as \reservoir~\cite{Reservoir_sampling_wiki}.
\reservoir aims to uniformly sample a pre-defined number of items from a stream, whose length is unknown in advance.
While \reservoir can be used as a primitive for different streaming algorithms, it can also be considered one as it single-passes through a stream with limited memory space.

\begin{table*}[!tb]
\centering
\caption{Taxonomy of tracker-based solutions (\S\ref{subsec:tracker-based}). 
Threat model \underline{\textbf{C}} denotes the RH \underline{characteristic} considered: basic (\(\Circle\)), Half-Double (\(\LEFTcircle\)), and RowPress (\(\CIRCLE\)).
\underline{\textbf{W-S}} shows whether the attacker is assumed to craft \underline{worst-case} access pattern regarding \underline{security} (\(\CIRCLE\)) or not (\(\Circle\)).
\underline{\textbf{A-P}} shows whether the attacker is assumed to craft \underline{adaptive} and \underline{performance} adversarial pattern (\(\CIRCLE\)), not adaptive (\(\LEFTcircle\)), not both (\(\Circle\)), or does not have a separate performance adversarial other than security worst-case (--).
For \underline{Related Algorithm} field, we mark * when the original study did not claim its correlation.
}
\label{tbl:tracker-based-defenses}
\Scale[0.9]{
\begin{tabular}{c|ccc|ccccc|c}
\Xhline{3\arrayrulewidth}
\multirow{2}{*}{\textbf{Goal}} & \multicolumn{3}{c|}{\textbf{Threat Model}} &
\multirow{2}{*}{\begin{tabular}[c]{@{}c@{}}\textbf{Related}\\\textbf{Algorithm}\end{tabular}} &
\multirow{2}{*}{\begin{tabular}[c]{@{}c@{}}\textbf{Mitigative}\\\textbf{Action}\end{tabular}} &
\multirow{2}{*}{\begin{tabular}[c]{@{}c@{}}\textbf{Security}\\\textbf{Guarantee}\end{tabular}} &
\multirow{2}{*}{\begin{tabular}[c]{@{}c@{}}\textbf{Process}\\\textbf{Tech.\,(Loc.)}\end{tabular}} &
\multirow{2}{*}{\begin{tabular}[c]{@{}c@{}}\textbf{Architectural}\\\textbf{Optimization}\end{tabular}} &
\multirow{2}{*}{\textbf{Reference}} \\ \cline{2-4}
 & \textbf{C} & \textbf{W-S} & \textbf{A-P} & & & & & & \\ 
\Xhline{3\arrayrulewidth}
% ---------- Low-cost tracker block ----------
\multirow{37}{*}{\begin{tabular}[c]{@{}c@{}}Low-cost tracker\end{tabular}}
& $\LEFTcircle$ & $\CIRCLE$ & -- & \misragries~\cite{scp-1982-misragries} & Refresh & Deterministic &
\begin{tabular}[c]{@{}c@{}}Logic\\(MC)\end{tabular} & NRR interface &
Graphene~\cite{micro-2020-graphene} \\ \cline{2-10}
& $\LEFTcircle$ & $\CIRCLE$ & $\LEFTcircle$ & \misragries~\cite{scp-1982-misragries} & Refresh & Deterministic &
\begin{tabular}[c]{@{}c@{}}Logic\\(MC)\end{tabular} &
\begin{tabular}[c]{@{}c@{}}Bank-level\\tracker sharing\end{tabular} &
ABACuS~\cite{security-2024-abacus} \\ \cline{2-10}
& $\CIRCLE$ & $\CIRCLE$ & $\LEFTcircle$ & \cmsketch~\cite{Algorithms-2005-countmin-sketch} & Refresh & Deterministic &
\begin{tabular}[c]{@{}c@{}}Logic\\(MC)\end{tabular} & Address caching &
COMET~\cite{hpca-2024-comet} \\ \cline{2-10}
& $\LEFTcircle$ & $\CIRCLE$ & $\LEFTcircle$ & *\cmsketch~\cite{Algorithms-2005-countmin-sketch} & Refresh & Deterministic &
\begin{tabular}[c]{@{}c@{}}Logic\\(MC)\end{tabular} & Tracker caching &
Hydra~\cite{isca-2022-hydra} \\ \cline{2-10}
& $\Circle$ & $\CIRCLE$ & $\Circle$ & \begin{tabular}[c]{@{}c@{}}*\cmsketch~\cite{Algorithms-2005-countmin-sketch},\\ *Sliding-Window~\cite{PODS-2004-approximate-counts-Arvind}\end{tabular} & Refresh & Deterministic &
\begin{tabular}[c]{@{}c@{}}Logic\\(MC)\end{tabular} & -- &
DBC~\cite{CAL-2019-DBC} \\ \cline{2-10}
& $\Circle$ & $\CIRCLE$ & -- & *\lossycounting~\cite{VLDB-2002-Lossy-sticky} & Refresh & Deterministic &
\begin{tabular}[c]{@{}c@{}}Logic\\(RCD)\end{tabular} &
\begin{tabular}[c]{@{}c@{}}Tracker set\\borrowing,\\ARR interface\end{tabular} &
TWiCe~\cite{isca-2019-twice} \\ \cline{2-10}
& $\Circle$ & $\CIRCLE$ & $\Circle$ & *Counter-trie~\cite{imc-2004-Counter-trie} & Refresh & Deterministic & \begin{tabular}[c]{@{}c@{}}Logic\\(RCD)\end{tabular} & Per-rank tracker &
CAT-TWO~\cite{access-2020-cat-two} \\ \cline{2-10}
& $\Circle$ & $\CIRCLE$ & $\Circle$ & *Counter-trie~\cite{imc-2004-Counter-trie} & Refresh & Deterministic &
\begin{tabular}[c]{@{}c@{}}Logic\\(MC)\end{tabular} & \begin{tabular}[c]{@{}c@{}}Access-pattern\\aware paramters\end{tabular} &
CBT~\cite{cal-2017-CBT,isca-2018-CBT} \\ \cline{2-10}
& $\LEFTcircle$ & $\CIRCLE$ & $\LEFTcircle$ & Exact Counting & Refresh & Deterministic & \begin{tabular}[c]{@{}c@{}}Logic\\(LLC)\end{tabular} &
\begin{tabular}[c]{@{}c@{}}LLC\\repurposing\end{tabular} &
START~\cite{hpac-2024-start} \\ \cline{2-10}
& $\Circle$ & $\CIRCLE$ & $\Circle$ & Exact Counting & Refresh & Deterministic & \begin{tabular}[c]{@{}c@{}}Logic\\(MC)\end{tabular} & Tracker caching &
CRA~\cite{cal-2015-CRA} \\ \cline{2-10}
& $\Circle$ & $\Circle$ & $\Circle$ & Exact Counting & \begin{tabular}[c]{@{}c@{}}Refresh\\(PRAC)\end{tabular} & \begin{tabular}[c]{@{}c@{}}Deterministic\\(bypassed)\end{tabular} &
\begin{tabular}[c]{@{}c@{}}Memory\\(DRAM)\end{tabular} & \begin{tabular}[c]{@{}c@{}}DRAM-based\\tracker\end{tabular} &
Panopticon~\cite{DRAMSec-2021-panopticon} \\ \cline{2-10}
& $\CIRCLE$ & $\Circle$ & -- & \begin{tabular}[c]{@{}c@{}}\misragries~\cite{scp-1982-misragries}\\(\spacesaving~\cite{ICDT-2005-Space-Saving})\end{tabular} & Refresh &
\begin{tabular}[c]{@{}c@{}}Empirical\\(bypassed)\end{tabular} &
\begin{tabular}[c]{@{}c@{}}Memory\\(DRAM)\end{tabular} & -- &
DSAC~\cite{arxiv-2023-dsac} \\ \cline{2-10}
& $\Circle$ & $\Circle$ & $\Circle$ & *Sticky-Sampling & Refresh &
\begin{tabular}[c]{@{}c@{}}Empirical\\(bypassed)\end{tabular} & \begin{tabular}[c]{@{}c@{}}Memory\\(DRAM)\end{tabular} & -- &
ProHIT~\cite{dac-2017-prohit} \\ \cline{2-10}
& $\Circle$ & $\Circle$ & $\Circle$ & *Reservoir-Sampling & Refresh &
\begin{tabular}[c]{@{}c@{}}Empirical\\(bypassed)\end{tabular} & \begin{tabular}[c]{@{}c@{}}Logic\\(MC)\end{tabular} & -- &
MRLoc~\cite{dac-2019-mrloc} \\ \cline{2-10}
& $\Circle$ & $\Circle$ & $\Circle$ & *\cmsketch~\cite{Algorithms-2005-countmin-sketch} & Refresh & Deterministic & \begin{tabular}[c]{@{}c@{}}Logic\\(MC)\end{tabular} & -- &
HammerFilter~\cite{ICCD-2021-hammerfilter} \\ \cline{2-10}
& $\LEFTcircle$ & $\Circle$ & $\Circle$ & *\stickysampling~\cite{VLDB-2002-Lossy-sticky} & Refresh & Empirical & \begin{tabular}[c]{@{}c@{}}Memory\\(DRAM)\end{tabular} & -- &
PROTEAS~\cite{arXiv-2024-proteas} \\ \cline{2-10}
& $\Circle$ & $\Circle$ & $\Circle$ & \begin{tabular}[c]{@{}c@{}}\emph{None}\\(rule-based)\end{tabular} & \begin{tabular}[c]{@{}c@{}}Refresh\\(RFM)\end{tabular} & Empirical & \begin{tabular}[c]{@{}c@{}}Logic\\(MC)\end{tabular} &
\begin{tabular}[c]{@{}c@{}}Anomaly detection,\\adaptive RFM\end{tabular} & MARC~\cite{arxiv-2025-marc} \\ \cline{2-10}
& $\Circle$ & $\Circle$ & $\Circle$ & \begin{tabular}[c]{@{}c@{}}\emph{None}\\(ML-based)\end{tabular} & Refresh & Empirical & \begin{tabular}[c]{@{}c@{}}Memory\\(DRAM)\end{tabular} &
\begin{tabular}[c]{@{}c@{}}ML-based\\anomaly detection\end{tabular} &
Joardar et al.~\cite{date-2022-learning} \\ \hline
%
% ---------- Low-cost tracker, Low-cost mitigation ----------
\multirow{2}{*}{
  \parbox[c][5\baselineskip][c]{2.0cm}{\centering
  Low-cost tracker, \\Low-cost\\mitigation}
}
& $\LEFTcircle$ & $\CIRCLE$ & $\LEFTcircle$ & \cmsketch~\cite{Algorithms-2005-countmin-sketch} & Throttle & Deterministic &
\begin{tabular}[c]{@{}c@{}}Logic\\(MC)\end{tabular} & Per-rank tracker &
BlockHammer~\cite{hpca-2021-blockhammer} \\ \cline{2-10}
& $\LEFTcircle$ & $\CIRCLE$ & -- & \misragries~\cite{scp-1982-misragries} & Shuffle &
\begin{tabular}[c]{@{}c@{}}Probabilistic\\(bypassed)\end{tabular} &
\begin{tabular}[c]{@{}c@{}}Logic\\(MC)\end{tabular} &
\begin{tabular}[c]{@{}c@{}}MIRAGE~\cite{usenix-2021-mirage}\\indirection table\end{tabular} &
RRS~\cite{asplos-2022-rrs} \\ \hline
% ---------- Low-cost tracker, Security improvement ----------
\begin{tabular}[c]{@{}c@{}}Low-cost tracker,\\Security\\improvement\end{tabular}
& $\LEFTcircle$ & $\CIRCLE$ & -- & \misragries~\cite{scp-1982-misragries} & Shuffle & Probabilistic & \begin{tabular}[c]{@{}c@{}}Logic\\(MC)\end{tabular} & LLC pinning &
SRS~\cite{hpca-2023-srs} \\ \hline
% ---------- Low-cost tracker, Queue-delay avoidance ----------
\multirow{2}{*}{
  \parbox[c][5\baselineskip][c]{2.0cm}{\centering
  Low-cost tracker, \\Queuing-delay\\avoidance}
}
& $\LEFTcircle$ & $\CIRCLE$ & -- & \misragries~\cite{scp-1982-misragries} & \begin{tabular}[c]{@{}c@{}}Refresh\\(RFM)\end{tabular} & Deterministic &
\begin{tabular}[c]{@{}c@{}}Memory\\(DRAM)\end{tabular} & Subarray parallelism &
ProTRR~\cite{sp-2022-protrr} \\ \cline{2-10}
& $\LEFTcircle$ & $\CIRCLE$ & -- & \begin{tabular}[c]{@{}c@{}}\misragries~\cite{scp-1982-misragries}\\(\spacesaving~\cite{ICDT-2005-Space-Saving})\end{tabular} & \begin{tabular}[c]{@{}c@{}}Refresh\\(RFM)\end{tabular} & Deterministic &
\begin{tabular}[c]{@{}c@{}}Memory\\(DRAM)\end{tabular} &
\begin{tabular}[c]{@{}c@{}}Adaptive RFM,\\No tracker reset\end{tabular} &
Mithril~\cite{hpca-2022-mithril} \\ \hline
% ---------- Queue-delay avoidance (two rows) ----------
\multirow{2}{*}{\begin{tabular}[c]{@{}c@{}}Queuing-delay\\avoidance\end{tabular}}
& $\CIRCLE$ & $\CIRCLE$ & $\LEFTcircle$ & Exact Counting & \begin{tabular}[c]{@{}c@{}}Refresh\\(PRAC)\end{tabular} & Deterministic & \begin{tabular}[c]{@{}c@{}}Memory\\(DRAM)\end{tabular} & \begin{tabular}[c]{@{}c@{}}DRAM-based\\tracker\end{tabular} &
MOAT~\cite{asplos-2025-moat} \\ \cline{2-10}
& $\LEFTcircle$ & $\CIRCLE$ & $\LEFTcircle$ & Exact Counting & \begin{tabular}[c]{@{}c@{}}Refresh\\(PRAC)\end{tabular} & Deterministic & \begin{tabular}[c]{@{}c@{}}Memory\\(DRAM)\end{tabular} & \begin{tabular}[c]{@{}c@{}}DRAM-based\\tracker\end{tabular} &
QPRAC~\cite{hpca-2025-qprac} \\ \hline
%
% ---------- Queue-delay avoidance, Improve PRAC ----------
\begin{tabular}[c]{@{}c@{}}Queuing-delay\\avoidance,\\Improve PRAC\\protocol\end{tabular}
& $\LEFTcircle$ & $\CIRCLE$ & $\LEFTcircle$ & Exact Counting & \begin{tabular}[c]{@{}c@{}}Refresh\\(PRAC)\end{tabular} & Deterministic & \begin{tabular}[c]{@{}c@{}}Memory\\(DRAM)\end{tabular} & \begin{tabular}[c]{@{}c@{}}DRAM-based\\tracker, modified\\PRAC\end{tabular} &
Chronus~\cite{hpca-2025-chronus} \\ \hline
%
% ---------- Low-cost tracker, Performance robustness ----------
\begin{tabular}[c]{@{}c@{}}Low-cost tracker,\\Performance\\robustness\end{tabular}
& $\LEFTcircle$ & $\CIRCLE$ & $\CIRCLE$ & \begin{tabular}[c]{@{}c@{}}*Adversarial-robust\\framework~\cite{JACM-2022-framework}\end{tabular} & Refresh & Probabilistic & \begin{tabular}[c]{@{}c@{}}Logic\\(MC)\end{tabular} & Lazy reset &
DAPPER~\cite{hpca-2025-dapper} \\ 
\Xhline{3\arrayrulewidth}
\end{tabular}
}
\end{table*}

\section{Taxonomy of Architectural RH Defenses}
\label{sec:3_taxonomy}

We present a taxonomy of RH defenses, leveraging streaming algorithms to establish a systematic classification.
We categorize architectural RH defense solutions into three types; i) tracker-based, ii) tracker-less, and iii) cooperative.
Tracker-based defenses utilize a form of hardware structures to identify aggressors and take mitigative action only when decided necessary.
Tracker-less defenses are oblivious to the activation pattern and takes continuous mitigative action to guarantee security.
Cooperative defenses are not a stand-alone defense, but aids other existing defenses when utilized together.
We analyze each RH defense solution using seven criteria: goal, threat model, related algorithm, mitigative action, security guarantee, process technology (location), and architectural optimization.

%-------------------------------------------------------------------------------
\subsection{Tracker-based Defenses}
\label{subsec:tracker-based}
%-------------------------------------------------------------------------------

Table~\ref{tbl:tracker-based-defenses} summarizes the taxonomy of tracker-based defenses.
The following criteria are considered:

\noindent
\textbf{1) Goal.}
\emph{Low-cost tracker} has been the goal of the majority of defense solutions.
The overhead of the tracker is mainly area and energy, which is often directly dependent on the target \rhth, tracker type, and the process technology.
We discuss these aspects in more detail (\S\ref{sec:4_good_combination}).
\emph{Low-cost mitigation} is also a goal of certain proposals, which introduced new forms of mitigative action other than refresh, such as throttle and shuffle.
\emph{Queuing-delay avoidance} and/or \emph{Improved RFM/PRAC protocol} is also the main focus of numerous proposals.
Such goals are especially crucial when building DRAM-side defense solutions, under the constraint of the MC-DRAM interface.
\emph{Performance robustness} can also be desired under the threat model where the defense solution is reverse-engineered.

\noindent
\textbf{2) Threat Model.}
We analyze the threat model of each defense with three different criteria: i) the level of considered RH \emph{characteristics} (\textbf{C}), ii) whether or not an activation pattern is the \emph{worst-case} in terms of \emph{security} (\textbf{W-S}), and iii) whether or not an activation pattern  is \emph{adaptive} and adversarial in terms of \emph{performance} (\textbf{A-P}).

(\textbf{C}) First, the threat model can consider various levels of RH characteristics: a rudimentary form~\cite{isca-2014-flipping} (\Circle), one that also considers Half-Double~\cite{sec-2022-halfdouble} (\LEFTcircle), and the one that also considers RowPress~\cite{isca-2023-rowpress} (\CIRCLE).

(\textbf{W-S}) Second, the threat model can assume that an attacker generates an arbitrary activation pattern, which can be worst-case in terms of security (\CIRCLE).
In other cases, the attacker is assumed to only generate realistic activation patterns affected by the system effects (\eg, MC scheduling~\cite{isca-2008-parbs}), or the worst-case pattern is unidentified (\Circle).

(\textbf{A-P}) Third, the threat model can assume an attacker that generates adaptive activation pattern that is performance adversarial (\CIRCLE).
Here, being adaptive means that the attacker can modify the activation pattern based on the reverse-engineered internal state of the RH defense.
This can occur if a mitigative action creates a timing side-channel~\cite{hpca-2025-dapper}, a vulnerability that adaptive attackers have exploited in other domains that use streaming algorithms~\cite{Algorithmica-1993-clocked,STOC-2013-robust}.
Such an adaptive attacker can be especially powerful when the tracker allows a form of tracker entry sharing (\eg, \cmsketch) because its randomness based on hashing can be nullified.
The threat model can also consider non-adaptive (\ie, not knowing the internal state of the defense) performance adversarial case (\LEFTcircle), or does not consider any performance adversarial case (\Circle).
We separately mark (--) when the defense solution does not have a separate performance adversarial activation pattern, other than the security worst-case pattern (\eg, \misragries).

\noindent
\textbf{3) Related Algorithm.}
Various forms of streaming algorithms are often crucial to the efficient trackers.
While some are utilized with clear intent (\eg, \cite{micro-2020-graphene}), it is also 
not uncommon for a defense solution to unintentionally reinvent an existing
algorithm (\eg, \cite{isca-2019-twice}).
We discuss such cases in more details later in \S\ref{sec:check_existing_algorithm}.
We use an asterisk (*) to indicate the cases where the original defense was unaware of the connection with the related algorithm.
A majority of the utilized algorithms are a form of frequency estimation or heavy-hitter algorithms: \misragries~\cite{scp-1982-misragries} (\spacesaving~\cite{ICDT-2005-Space-Saving}), \cmsketch~\cite{Algorithms-2005-countmin-sketch}, \lossycounting~\cite{VLDB-2002-Lossy-sticky}, and \stickysampling~\cite{VLDB-2002-Lossy-sticky}.
There are also other relevant algorithms such as \reservoir~\cite{JASA-1962-Method-4}, sliding-window algorithm~\cite{PODS-2004-approximate-counts-Arvind}, and adversarial-robust framework~\cite{JACM-2022-framework}.
Several tracker-based defenses employ a rudimentary form of tracking; exact counting.
Such works focus on architectural optimizations or low-cost mitigation mechanisms.

There are two cases~\cite{arxiv-2025-marc,date-2022-learning} not utilizing a form of streaming algorithms, to the best of our knowledge.
Both leverage a pure form of anomaly detection, of rule-based~\cite{arxiv-2025-marc} or machine learning-based~\cite{date-2022-learning}.
However, both demonstrated only an empirical security against tested attack patterns.
We denote these two cases as \emph{None} in Table~\ref{tbl:tracker-based-defenses}.

\begin{table*}[!tb]
\centering
\caption{Taxonomy of tracker-less solutions (\S\ref{subsec:tracker-less}), which follows the same notation with Table~\ref{tbl:tracker-based-defenses}.
}
\label{tbl:tracker-less-defenses}
\vspace{-0.05in}
\Scale[0.9]{
\begin{tabular}{c|ccc|ccccc|c}
\Xhline{3\arrayrulewidth}
\multirow{2}{*}{\textbf{Goal}} & \multicolumn{3}{c|}{\textbf{Threat Model}} &
\multirow{2}{*}{\begin{tabular}[c]{@{}c@{}}\textbf{Related}\\\textbf{Algorithm}\end{tabular}} &
\multirow{2}{*}{\begin{tabular}[c]{@{}c@{}}\textbf{Mitigative}\\\textbf{Action}\end{tabular}} &
\multirow{2}{*}{\begin{tabular}[c]{@{}c@{}}\textbf{Security}\\\textbf{Guarantee}\end{tabular}} &
\multirow{2}{*}{\begin{tabular}[c]{@{}c@{}}\textbf{Process}\\\textbf{Tech.\,(Loc.)}\end{tabular}} &
\multirow{2}{*}{\begin{tabular}[c]{@{}c@{}}\textbf{Architectural}\\\textbf{Optimization}\end{tabular}} &
\multirow{2}{*}{\textbf{Reference}} \\ \cline{2-4}
 & \textbf{C} & \textbf{W-S} & \textbf{A-P} & & & & & & \\ 
\Xhline{3\arrayrulewidth}
% ---------- Low-cost tracker block ----------
\begin{tabular}[c]{@{}c@{}}Sampling-based\\defense\end{tabular} & $\Circle$ & $\CIRCLE$ & -- & Bernoulli-sampling & Refresh & Probabilistic &
\begin{tabular}[c]{@{}c@{}}Logic\\(MC)\end{tabular} &
-- &
PARA~\cite{isca-2014-flipping} \\ \hline
\multirow{7}{*}{\begin{tabular}[c]{@{}c@{}}DRAM-side\\sampling-based\\defense\end{tabular}}
& $\Circle$ & $\CIRCLE$ & -- & *\reservoir~\cite{JASA-1962-Method-4} & Refresh & Probabilistic &
\begin{tabular}[c]{@{}c@{}}Memory\\(DRAM)\end{tabular} & -- &
PARFM~\cite{hpca-2022-mithril} \\ \cline{2-10}
& $\LEFTcircle$ & $\CIRCLE$ & -- & *\reservoir~\cite{JASA-1962-Method-4} & Refresh & Probabilistic &
\begin{tabular}[c]{@{}c@{}}Memory\\(DRAM)\end{tabular} & -- &
PrIDE~\cite{isca-2024-pride} \\ \cline{2-10}
& $\LEFTcircle$ & $\CIRCLE$ & -- & *\reservoir~\cite{JASA-1962-Method-4} & Refresh & Probabilistic &
\begin{tabular}[c]{@{}c@{}}Memory\\(DRAM)\end{tabular} & \begin{tabular}[c]{@{}c@{}}postponed\\refresh queue\end{tabular} &
MINT~\cite{MICRO-2024-mint} \\ \hline
\multirow{3}{*}{\begin{tabular}[c]{@{}c@{}}Low-cost\\mitigation\end{tabular}} 
& $\LEFTcircle$ & $\CIRCLE$ & -- & \begin{tabular}[c]{@{}c@{}}\emph{None}\\(Oblivious)\end{tabular} & Refresh & Deterministic &
\begin{tabular}[c]{@{}c@{}}Memory\\(DRAM)\end{tabular} & \begin{tabular}[c]{@{}c@{}}Additional\\sense amplifier\end{tabular} &
REGA~\cite{sp-2023-rega} \\ \cline{2-10}
& $\LEFTcircle$ & $\CIRCLE$ & -- & \begin{tabular}[c]{@{}c@{}}\emph{None}\\(Oblivious)\end{tabular} & Shuffle & Probabilistic &
\begin{tabular}[c]{@{}c@{}}Memory\\(DRAM)\end{tabular} & \begin{tabular}[c]{@{}c@{}}Remapping row,\\subarray pairing\end{tabular} &
SHADOW~\cite{hpca-2023-shadow} \\
\Xhline{3\arrayrulewidth}
\end{tabular}
}
\end{table*}

\noindent
\textbf{4) Mitigative Action.}
The defense solution can leverage one of three mitigative actions; i) refresh, ii) throttle, and iii) shuffle.
Refresh means the RH victim rows are additionally refreshed other than the ordinary auto-refresh, before an RH bitflip occurs, and can be realized through interfaces such as TRR, RFM, and PRAC.
While all belong to the refresh category, they differ in where and when the refresh is triggered.
From a streaming algorithm perspective, these mechanisms correspond to periodically bounding or resetting element frequencies, analogous to leaky-bucket or sliding-window~\cite{PODS-2004-approximate-counts-Arvind} models.
Throttle means the host limits the rate of activation that an aggressor can achieve.
Shuffle means the aggressor/victim row location is swapped with a random location.
The relationship between the mitigative action and streaming algorithms are discussed in \S\ref{subsec:4_mitigative_action}.

\noindent
\textbf{5) Security Guarantee.}
There are three types of security guarantees that RH defense solutions provide; i) deterministic, ii) probabilistic, and iii) empirical.
\emph{Deterministic} guarantee means that the solution provides a security guarantee that always holds for a given \rhth.
\emph{Probabilistic} guarantee means that the solution provides a security guarantee that holds with high probability, yet with non-zero probability of failure.
Both deterministic and probabilistic safety guarantees accompany a form of mathematical proof with regard to the worst-case activation pattern against security.
\emph{Empirical} guarantee means that it is tested safe against a set of crafted attack patterns.
Such attack patterns can be either hand-crafted, or based on a form of fuzzer~\cite{sp-2022-blacksmith}.

\noindent
\textbf{6) Process Technology (Location).}
The defense solution can reside on the host-side or DRAM-side, which determines what type of process technology it can utilize.
The large differences between logic and memory process technologies become more pronounced in the tracker-based defenses, as the area overhead of a tracker is bloated when using the memory process technology.
Also, DRAM-side defense solutions require a form of interface to allow timing slack for mitigative action, such as TRR, ARR~\cite{isca-2019-twice}, NRR~\cite{micro-2020-graphene}, RFM~\cite{jedec-2017-ddr4}, and PRAC~\cite{jedec-2024-ddr5}.

\noindent
\textbf{7) Architectural Optimization.}
Various architectural techniques are utilized for each RH defense.
We focus on the techniques that are orthogonal to the tracker or related algorithms themselves, such as new interfaces (\eg, NRR), caching, or DRAM cell-based tracking.

%-------------------------------------------------------------------------------
\subsection{Tracker-less Defenses}
\label{subsec:tracker-less}
%-------------------------------------------------------------------------------

Table~\ref{tbl:tracker-less-defenses} summarizes the taxonomy of tracker-less defenses.
The same criteria as for the tracker-based defense are adopted.
We omit the discussion of mitigative action, security guarantee, and process technology (location), and architectural optimization.

\noindent
\textbf{1) Goal.}
\emph{Sampling-based defense} is adopted by multiple prior studies.
By continuously and randomly selecting an activated row and refreshing its victim rows, the defense solution provides probabilistic safety guarantee.
The probability of safety guarantee is defined by the sampling-rate.
When the rate is higher, the attacker is less likely to evade random selection and refresh.
\emph{DRAM-side sampling-based defense} has been the goal of multiple prior studies, which is more tricky due to the MC-DRAM interface.
\emph{Improved RFM protocol} as a main goal is in line with such an approach.
\emph{Low-cost mitigation}, another goal of prior art, aims to mitigate the performance overhead of defense solutions.

\noindent
\textbf{2) Threat Model.}
Because tracker-less defenses are agnostic to the activation pattern, there is no particular activation pattern that is performance adversarial.

\noindent
\textbf{3) Related Algorithm.}
DRAM-side sampling-based defenses solve the same problem that \reservoir~\cite{JASA-1962-Method-4} solves, although no prior study has identified this connection.
We discuss this in further details later in \S\ref{sec:check_existing_algorithm}.
Two cases~\cite{sp-2023-rega,hpca-2023-shadow} are not directly related to streaming algorithms.
Both methods employ DRAM microarchitecture modifications, executing mitigation \emph{obliviously} at every activation without requiring tracking or sampling.
These two approaches can be viewed as a new form of mitigation, similar to TRR, RFM, and PRAC.

%-------------------------------------------------------------------------------
\subsection{Cooperative Defenses}
\label{subsec:cooperative}
%-------------------------------------------------------------------------------

Table~\ref{tbl:cooperative-defenses} presents our taxonomy of cooperative defenses and frameworks.
We only discuss their goals in this category, due to their diverse scope and form.
Defense solutions in this category do not provide a stand-alone protection guarantee.
Some studies provide \emph{Performance attack mitigation}, while the RH security guarantee is provided by the co-running defense solution.
A body of works improves the \emph{Efficiency} of other RH defense solutions.
Others help \emph{BER (bit-error-rate) reduction} caused by RH bitflips.
Lastly, a group of proposals keeps the \emph{Integrity} of data even in the case of RH-induced uncorrectable errors, preventing RH from becoming a critical attack vector.
We consider such works as also cooperative, as they by themselves do not prevent RH bitflips or resulting reliability issues or DoS attacks.

\begin{table*}[!tb]
\centering
\caption{Taxonomy of Cooperative solutions and frameworks (\S\ref{subsec:cooperative}).
}
\label{tbl:cooperative-defenses}
\Scale[0.95]{
\renewcommand{\arraystretch}{1.3}
\begin{tabular}{c|l|c}
\Xhline{3\arrayrulewidth}
\textbf{Goal} & \multicolumn{1}{c|}{\textbf{Focus}} & \textbf{Reference} \\
\Xhline{3\arrayrulewidth}
\begin{tabular}[c]{@{}c@{}}Performance attack mitigation\end{tabular} & Throttle aggressor rows to prevent performance attack & BreakHammer~\cite{micro-2024-breakhammer} \\ \hline
Efficiency & Allow other defenses to efficiently handle Half-Double~\cite{sec-2022-halfdouble} attack & AQUA~\cite{micro-2022-aqua} \\ \hline
Efficiency & Allow other defenses to transparently execute mitigative refresh & HiRA~\cite{micro-2022-hira} \\ \hline
Efficiency & Allow other defenses to exploit spatial variation of RH vulnerability & Sv\"ard~\cite{hpca-2024-spatial} \\ \hline
Efficiency & Decrease the hot rows by obfuscate cacheline to DRAM row mapping & Rubix~\cite{asplos-2024-rubix} \\ \hline
Efficiency & Allow other defenses to unify RowPress and RowHammer mitigation & Impress~\cite{micro-2024-impress} \\ \hline
Efficiency & Improve security of other defenses with address remapping and ECC & Cube~\cite{micro-2023-cube} \\ \hline
Efficiency & Improve the efficiency of RFM mechanism & AutoRFM~\cite{hpca-2025-autorfm} \\ \hline
BER reduction & Reduce RH-induced BER with address remapping and ECC & Kim et al.~\cite{TC-2019-Effective} \\ \hline
BER reduction & Reduce RH-induced BER with address remapping and ECC & Wang et al.~\cite{ICCD-2019-reinforce} \\ \hline
BER reduction & Reduce RH-induced BER with address remapping and ECC & RAMPART~\cite{memsys-2023-rampart} \\ \hline
Integrity & Detect RH bitflips using authentication codes & CSI~\cite{sp-2023-csi} \\ \hline
Integrity & Detect RH bitflips using authentication codes & SafeGuard~\cite{hpca-2022-safeguard} \\ \hline
Integrity & Detect RH bitflis on PageTable using authentication codes & PT-Guard~\cite{dsn-2023-pt-guard} \\
\Xhline{3\arrayrulewidth}
\end{tabular}
}
\end{table*}

%-------------------------------------------------------------------------------
\section{Which Algorithm to Use When?}
%-------------------------------------------------------------------------------
\label{sec:4_good_combination}

Choosing the appropriate streaming algorithm for a given situation is crucial; \rhth, process technology, location, type of mitigative action, and architectural optimization all affect the optimal choice.
This section discusses the details of the streaming algorithms RH defense solutions can utilize.
First, we compare the overheads of various algorithms, focused on the area overhead of the frequency estimation algorithms for tracker-based defenses (\S\ref{subsec:4_area_overhead}).
We focus on the mathematically derivable aspect, not the application-specific and configurable components.
Second, we discuss what mitigative action fits well with the utilized algorithms, regarding the type of error bounds they provide (\S\ref{subsec:4_mitigative_action}).
Last, we discuss what architectural techniques can be utilized to aid the underlying algorithms  (\S\ref{subsec:4_architectural_optimization}).

\begin{table}[!tb]
    \centering
    \small
    \caption{Symbols and terminologies in \S\ref{sec:4_good_combination}}
    \label{tbl:symbols_streaming}
    \vspace{-0.05in}
    \Scale[0.95]{
    \begin{tabular}{p{0.13\columnwidth} p{0.82\columnwidth}} 
        \Xhline{3\arrayrulewidth}
        \textbf{Symbol} & \textbf{Description} \\ 
        \Xhline{1.5\arrayrulewidth}
        %$N$      & Total stream length (\ie, number of activations within tREFW window). Maximum value ($\frac{32ms-8192\times 350ns}{48ns}$), unless stated otherwise. \\
        \fest    & Algorithm's estimated frequency (\ie, the number of activations) of an element (\ie, a row address). \\
        \freal & Real frequency of an element. \\
        $C\{\}$   & Counter structure of the algorithm, with tuples of address and \fest. \\
        $N$     & Length of a stream. \\
        $\epsilon$  & Error rate parameter. $\epsilon N$ defines the actual error. \\
        $\delta$    & Bound confidence parameter. \\
        \Xhline{0.5\arrayrulewidth}
        \Xhline{3\arrayrulewidth}
    \end{tabular}
    }
    \vspace{-0.03in}
\end{table}

\begin{table*}[!tb]
\centering
\caption{Summarization of representative algorithms and related RowHammer solutions.}
\label{tbl:algorithms}
\Scale[0.9]{
\newcolumntype{P}[1]{>{\centering\arraybackslash}p{#1}}
\begin{tabular}{c|P{4cm}P{2.5cm}|P{2.5cm}P{2.5cm}P{2.5cm}}
\Xhline{3\arrayrulewidth}
\hline
  \textbf{Algorithm} & \textbf{Error Bound} & \begin{tabular}[c]{@{}c@{}}\textbf{Space}\\\textbf{Complexity}\end{tabular} & \begin{tabular}[c]{@{}c@{}} \textbf{Related} \\\textbf{RH Solution}\end{tabular} & \begin{tabular}[c]{@{}c@{}}\textbf{Required}\\\textbf{Hardware}\end{tabular} & \begin{tabular}[c]{@{}c@{}}\textbf{Technology}\\\textbf{(Location)}\end{tabular} \\ 
\hline
\Xhline{3\arrayrulewidth}
\multirow{3}{*}{\begin{tabular}[c]{@{}c@{}}\textbf{\spacesaving \&}\\ \textbf{\misragries~\cite{scp-1982-misragries,ICDT-2005-Space-Saving}}\end{tabular}} 
    & \multirow{3}{*}{\begin{tabular}[c]{@{}c@{}}$\frealeq\leq\festeq\leq\frealeq+\epsilon N$ \&\\ $\frealeq-\epsilon N\leq\festeq\leq\frealeq$\end{tabular}} 
    & \multirow{3}{*}{$\frac{1}{\epsilon}(logN)$} 
    & \begin{tabular}[c]{@{}c@{}}Graphene~\cite{micro-2020-graphene}, RRS~\cite{asplos-2022-rrs}\\AQUA~\cite{micro-2022-aqua}, SRS~\cite{hpca-2023-srs}\end{tabular} 
    & \begin{tabular}[c]{@{}c@{}}CAM or\\Linked-list+SRAM~\cite{INFOCOM-2016-Space-saving-linked-list}\end{tabular} 
    & \begin{tabular}[c]{@{}c@{}}Logic Process\\(MC)\end{tabular} \\
\cline{4-6}
    &  
    &  
    & \begin{tabular}[c]{@{}c@{}}Mithril~\cite{hpca-2022-mithril}\\ProTRR~\cite{sp-2022-protrr}\end{tabular}
    & CAM 
    & \begin{tabular}[c]{@{}c@{}}Memory Process\\(DRAM-side)\end{tabular} \\ 
\hline
\textbf{\lossycounting~\cite{VLDB-2002-Lossy-sticky}} 
    & $\frealeq-\epsilon N\leq\festeq\leq\frealeq$ 
    & $\frac{1}{\epsilon}log(\epsilon N)log(N)$ 
    & TWiCe~\cite{isca-2019-twice} 
    & CAM+SRAM 
    & \begin{tabular}[c]{@{}c@{}}Logic Process\\(SPD chip)\end{tabular} \\ 
\hline
\textbf{CountMin-Sketch~\cite{Algorithms-2005-countmin-sketch}} 
    & $\frealeq\leq\festeq\overset{1-\delta}{\leq}\frealeq+\epsilon\|f_{-a}\|_1$
    & $\frac{1}{\epsilon}log\frac{1}{\delta}logN$ 
    & \begin{tabular}[c]{@{}c@{}}BlockHammer~\cite{hpca-2021-blockhammer}\\CoMet~\cite{hpca-2024-comet}, Hydra~\cite{isca-2022-hydra}\end{tabular} 
    & SRAM 
    & \begin{tabular}[c]{@{}c@{}}Logic Process\\(MC)\end{tabular} \\ 
\hline
\textbf{Count-Sketch~\cite{icalp-2002-count-sketch}} & $\frealeq-\epsilon\|f_{-i}\|_2\overset{1-\delta}{\leq}\festeq\overset{1-\delta}{\leq}\frealeq+\epsilon\|f_{-i}\|_2
$ & $\frac{1}{\epsilon^2}log\frac{1}{\delta}log{N}$ & -- & -- & -- \\ 
\hline
\textbf{\stickysampling~\cite{VLDB-2002-Lossy-sticky}} & $\frealeq-\epsilon N\overset{1-\delta}{\leq}\festeq\leq\frealeq$ & $\frac{2}{\epsilon}log(\frac{1}{2\epsilon\delta})logN$$\dagger$ & -- & -- & -- \\ 
\hline
\begin{tabular}[c]{@{}c@{}}\textbf{Probabilistic}\\\textbf{Wavelet Synopsis~\cite{tods-2004-wavelet}}\end{tabular} & $\frealeq-\epsilon N\overset{1-\delta}{\leq}\festeq\overset{1-\delta}{\leq}\frealeq+\epsilon N$ & $\frac{1}{\epsilon^2}log\frac{1}{\delta}logN$ & -- & -- & -- \\ 
\hline
\textbf{Counter-trie~\cite{imc-2004-Counter-trie}}
    & -- 
    & -- 
    & \begin{tabular}[c]{@{}c@{}}Counter-based \\Tree~\cite{cal-2017-CBT,isca-2018-CBT}$\ddagger$\end{tabular} 
    & SRAM 
    & \begin{tabular}[c]{@{}c@{}}Logic Process\\(MC)\end{tabular} \\
\hline
\Xhline{3\arrayrulewidth}
\multicolumn{6}{c}{\begin{tabular}[c]{@{}l@{}}$\|f_{-i}\|_1$ \& $\|f_{-i}\|_2$ refer to $L1$ \& $L2$ norm of the total stream, excluding the element $i$ when the error bound is concerning element $i$.\\$\dagger$ Assuming that support parameter ($s$) is $2\epsilon$. $\ddagger$ Large similarities in retrospect.\end{tabular}
 }
\end{tabular}
}
\end{table*}

%-------------------------------------------------------------------------------
\subsection{Symbol and Terminology Definition}
\label{subsec:4_symbol_term}
%-------------------------------------------------------------------------------

We use the following symbols and terminologies in discussing and comparing various streaming algorithms that can be utilized for frequency estimation or heavy-hitter problems.
Table~\ref{tbl:symbols_streaming} summarizes the symbols and terminologies used henceforth.
The total length of the stream is $N$.
In our setting, the maximum possible $N$ is the maximum number of activations within a tREFW window (\eg, $\frac{32ms-8192\times 350ns}{48ns}$).
We consider $N$ to be the maximum unless stated otherwise.

For a certain element (\ie, row address), its estimated or reported frequency is denoted as \fest.
Its true frequency, which can only be known if one keeps exact activation counts of all row addresses, is denoted as \freal.
For each algorithm, the relationship between \fest and \freal is defined by certain inequalities, dictated by various parameters.
First, the error parameter ($\epsilon$) affects the accuracy of the algorithm regarding the total stream length $N$.
For example, the \spacesaving algorithm provides an upper- and lower-bound of \fest regarding \freal and $\epsilon N$.
Typically, we set $\epsilon$ such that $\epsilon N = \rhtheq$.
Second, the bound confidence parameter ($\delta$) defines the probability that a bound holds for certain probabilistic algorithms.
For example, the \fest upper-bound of CountMin-Sketch fails with probability $\delta$.
Such parameters also affect the space complexity of each algorithm.

Table~\ref{tbl:algorithms} summarizes the \fest bounds and space complexities of relevant streaming algorithms that can be utilized for heavy-hitter or frequency-estimation problems; \misragries/\spacesaving~\cite{scp-1982-misragries,ICDT-2005-Space-Saving}, \lossycounting~\cite{VLDB-2002-Lossy-sticky}, \cmsketch~\cite{Algorithms-2005-countmin-sketch}, \csketch~\cite{icalp-2002-count-sketch}, \stickysampling~\cite{VLDB-2002-Lossy-sticky}, Probabilistic Wavelet Synopsis~\cite{tods-2004-wavelet}, and Counter-trie~\cite{imc-2004-Counter-trie}.
Some algorithms have been utilized in prior works of literature, although some have, in retrospect, rediscovered the algorithm to a certain extent.

%-------------------------------------------------------------------------------
\subsection{Area Overhead of Tracker-based Defenses: Impact of Technology and RH Threshold}
\label{subsec:4_area_overhead}
%-------------------------------------------------------------------------------

\begin{table}[tb!]
    \centering
    \small
    \caption{Area of SRAM, CAM, and DRAM, by the logic or memory process technology~\cite{process-logic, process-dram, circuits-2016-CAM-area, taco-2017-cacti, hcs-2019-process_scaling}.
    }
    \label{tbl:area_overhead}
    \Scale[0.9]{
    \begin{tabular}{p{0.22\columnwidth} p{0.28\columnwidth} p{0.4\columnwidth}} 
        \Xhline{3\arrayrulewidth}
        \textbf{Hardware} & \textbf{Technology} & \textbf{Area overhead (\umsquare/bit)} \\
        \Xhline{1.5\arrayrulewidth}
        \multirow{2}{*}{SRAM} & Logic & \textcolor{white}{0}0.0263 \\
                            & Memory & \textcolor{white}{0}7.3 \\
        \Xhline{1.5\arrayrulewidth}
        \multirow{2}{*}{CAM} & Logic & \textcolor{white}{0}0.0526 \\
                            & Memory & 14.6 \\
        \Xhline{1.5\arrayrulewidth}
        DRAM & Memory & \textcolor{white}{0}0.00317 \\
        \Xhline{3\arrayrulewidth}
    \end{tabular}
    }
    \vspace{-0.08in}
\end{table}

\begin{figure*}[tb!]
  \center
  \vspace{0.0in}
  \includegraphics[width=0.91\linewidth]{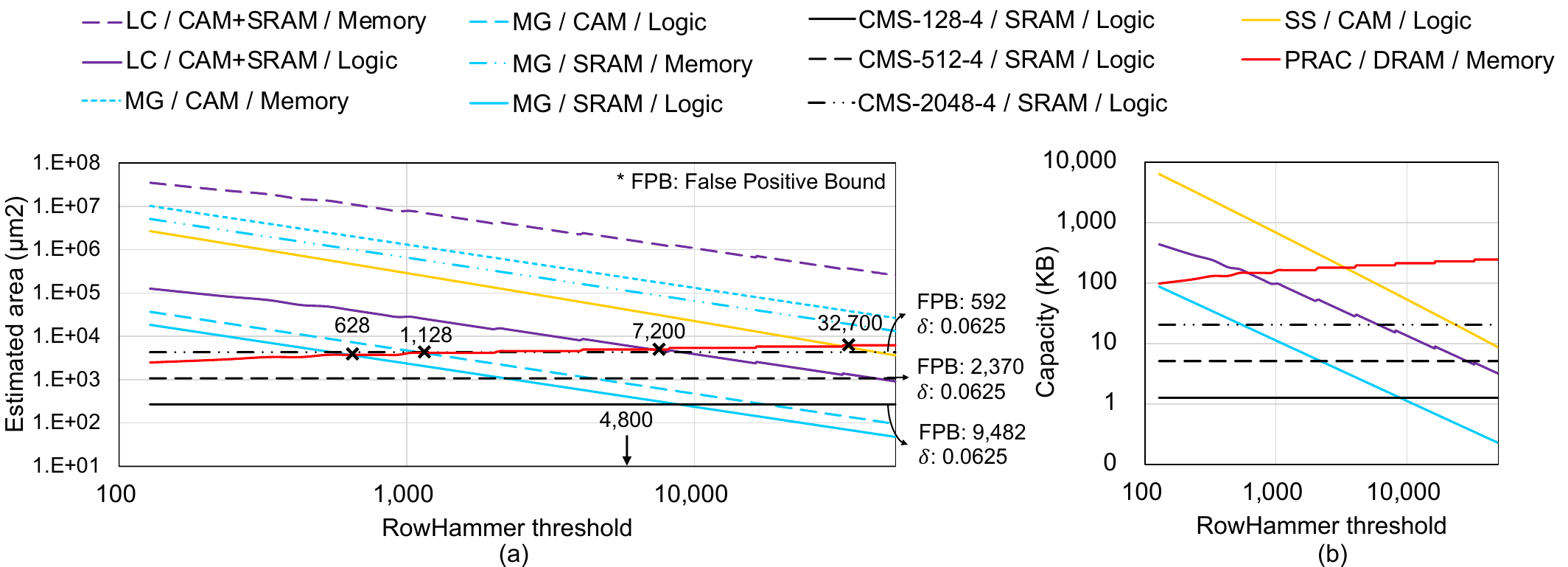}
  \vspace{-0.08in}
  \caption{(a) Per bank area overhead of various algorithms, with consideration of storage hardware and process technology. (b) Per bank capacity overhead of various algorithms. LC, MG, SS, CMS, and PRAC stand for \lossycounting, \misragries, \stickysampling, CountMin-Sketch, and Per Row Activation Count, respectively. 
  The notation, \eg, PRAC / DRAM / Memory, specifies the algorithm / hardware / process technology.
  The notation, \eg, CMS-128-4, denotes the \cmsketch with specified row width (\eg, 128) and the number of rows/hashes (\eg, 4).
  }
  \vspace{-0.04in}
  \label{fig:area_algorithms}
\end{figure*}

We compare the necessary area overhead of the streaming algorithms that are utilized for various defense solutions (Figure~\ref{fig:area_algorithms}).
We focus on the algorithmic perspective that is mathematically defined by security guarantee, instead of factors that are used for performance and affected by application behavior.
We omit the area overhead of additional hardware other than the frequency estimation tracker.
This is because their size is either empirically determined based on simulation~\cite{isca-2022-hydra,hpca-2024-comet}, or is constant across a changing \rhth~\cite{asplos-2022-rrs,hpca-2023-srs}.
We omit \csketch and Probabilistic Wavelet Synopsis, as their area overheads explode when \rhth drops. 
This is because there is a $\frac{1}{\epsilon^2}$ term in their space complexities.
For comparison, we also include DRAM cell-based per-row counters~\cite{DRAMSec-2021-panopticon}.

\noindent
\textbf{Methodology.}
First, we cannot directly apply \rhth to match $\epsilon N$.
Prior studies often adopt a periodic reset of the counter structure~\cite{isca-2019-twice,micro-2020-graphene,hpca-2021-blockhammer,sp-2022-protrr,asplos-2022-rrs,hpca-2023-srs,hpca-2024-comet}.
Because this reset can be asynchronous compared to the auto-refresh of each row, the target threshold must be $\rhtheq/2$ to be conservative.
Also, the fact that a victim can receive hammering from two adjacent aggressor rows forces additional halving.
Thus, we set $\epsilon N$ to match $\rhtheq/4$.
PRAC is an exception and is applied with $\rhtheq/2$, as it does not require a reset.

Second, we consider what storage types and process technologies each algorithm can utilize.
Simply referring to the space complexities of each algorithm does not provide a complete picture of the area overhead (compare Figure~\ref{fig:area_algorithms}(a) with \ref{fig:area_algorithms}(b))
as each algorithm requires different storage types, and can utilize different process technologies depending on their location: processor- or DRAM-side.
Table~\ref{tbl:area_overhead} summarizes our assumptions for the area overhead of SRAM, CAM, and DRAM (\umsquare/bit).
We base our calculation on the cell efficiency of SRAM on the logic process~\cite{process-logic} and DRAM in a memory process~\cite{process-dram}.
We assume that a CAM cell occupies twice the area of an SRAM cell~\cite{circuits-2016-CAM-area, taco-2017-cacti}.
%
%CAM is assumed to be double the area of SRAM~\cite{circuits-2016-CAM-area, taco-2017-cacti}.
%
Finally, we conservatively scale the sizes of SRAM and CAM in the memory process by a factor of 10~\cite{hcs-2019-process_scaling}.

\noindent
\textbf{Observation-i).}
\stickysampling and \lossycounting require a higher area and capacity overhead compared to other techniques.
In the case of \stickysampling, it can handle different queries other than frequency estimation such as quantile or distinct element; thus, it holds more information compared to the optimal \misragries/\spacesaving.
Both \stickysampling and \lossycounting can have dynamically changing storage overhead according to the actually encountered access pattern.
While this can be effective in software when handling \averagecase, this does not fit the RH-solution context, which builds a fixed hardware considering \worstcase.

\noindent
\textbf{Observation-ii).}
\misragries/\spacesaving requires a small area and capacity overhead in general cases, especially as \rhth is relatively high.
Also, when comparing only the capacity overhead, it shows a smaller overhead even when compared to PRAC.
In the case of logic process-based SRAM/CAM, \misragries/\spacesaving requires an order of magnitude smaller area overhead compared to PRAC when \rhth is 4,800 (492/984 and 4,985, respectively).
4,800 is the lowest observed \rhth value~\cite{isca-2020-revisiting}.
However, when \misragries/\spacesaving utilizes memory process-based SRAM/CAM, they exhibit an order of magnitude higher area overhead compared to PRAC (136,670/273,341, respectively).
Also, when \rhth drops below an extreme scale of 628, even the logic process-based \misragries/\spacesaving incurs higher area overhead compared to PRAC.

\noindent
\textbf{Observation-iii).}
Certain \cmsketch configurations provide a superior area overhead, even when compared to PRAC on all \rhth.
This is because the capacity and area overheads of \cmsketch 
(which determine its ability to meet the \fest lower-bound shown in Table~\ref{tbl:algorithms}) do not inherently depend on the target \rhth in the same way as other algorithms.
However, the upper-bound, which dictates the false-positive, is affected by $\epsilon$ and the total length of the stream, or $N$.
The width of \cmsketch is defined by $\epsilon$; $\epsilon = 2/width$.
Also, the bound confidence $\delta$ of the false positive defines the required number of hashes/rows; $\delta = 1/(2^{\#rows})$.
Assuming the worst case of the maximum length $N$ (\ie, $\frac{tREFW-tREFW/tREFI\times tRFC}{tRC}$), the false positive bound and its confidence are denoted at the right side of Figure~\ref{fig:area_algorithms}(a).
The false-positive bound of the largest evaluated \cmsketch with 2048 width 4 rows is 592, with a confidence of 0.0625.
This can be too wide (and thus weak) a bound when the \rhth drops to an extreme scale.

Despite such characteristics, false positives affect only performance, not security.
Thus, prior studies such as BlockHammer~\cite{hpca-2021-blockhammer} or CoMeT~\cite{hpca-2024-comet} empirically chose the \cmsketch size based on the evaluation of the \averagecase.
The Recent Aggressor Table (RAT) of CoMeT~\cite{hpca-2024-comet} does not mitigate the possibility of such false-positive cases, but rather solves the problem of inability to decrement counters after post-mitigation.

\noindent
\textbf{Observation-iv).}
PRAC can require a smaller area overhead compared to other techniques, especially when \rhth decreases.
First, it is because only PRAC can utilize the DRAM cell as storage, which has orders of magnitude smaller area overhead compared to SRAM and CAM (Table~\ref{tbl:area_overhead}).
Second, each entry of PRAC is narrower compared to other algorithms because it does not require row address bits.
Finally, as \rhth decreases, PRAC can reduce the counter bit-width and area overhead, whereas other algorithms experience a superlinear increase in area overhead.
Given that the area advantage of PRAC primarily stems from using smaller DRAM cells, future solutions employing DRAM cells for streaming algorithm-based trackers may also prove valuable.

%-------------------------------------------------------------------------------
\subsection{Algorithmic Requirements for Different Mitigative Action}
\label{subsec:4_mitigative_action}
%-------------------------------------------------------------------------------

The efficacy of adopting a streaming algorithm is also dependent on what mitigation techniques one is willing to use; refresh, throttle, and shuffle.

\noindent
\textbf{i) Refresh.}
Refresh as mitigative action causes \freal (activation count) of the aggressor row to become zero.
In the case of the tracker-based solutions, the count value of such a row should be decremented, if not reset to zero.
Without such a proper post-mitigation counter update, it can cause continuous mitigation after one surpasses a certain threshold, or can waste mitigation opportunities by repeatedly selecting an already mitigated row.

This is a unique challenge not typically found in conventional streaming algorithm applications.
Traditional contexts such as the (strict) turnstile model~\cite{FTTCS-2005-muthukrishnan-textbook} consider a case where not only insertion, but also \emph{deletion} of an item can occur.
However, this does not apply to our case, as we do not even know \emph{how much} we want to delete as we do not know the \freal of any element prior to the mitigative refresh.

The second best we get is the deterministic upper-bound of the \fest.
%when \freal becomes zero after mitigative refresh.
%
For example, \lossycounting and \misragries has the \freal upper-bound up until \freal (Table~\ref{tbl:algorithms}); thus we can safely reset the related count to zero.
The upper-bound in \spacesaving is until $\frealeq+\epsilon N$; thus, we can decrease to as low as $\epsilon N$ (\ie, \minentry).
Such a property was recognized and utilized in ProTRR~\cite{sp-2022-protrr} and Mithril~\cite{hpca-2022-mithril}.
In the case of Graphene~\cite{micro-2020-graphene}, the post-mitigation update was circumvented by only triggering a refresh at the integer multiple of a predefined threshold value.
This is based on the observation that when such a threshold is properly set, any element that triggers mitigation at least once, never gets swapped by a different element.

In the case of \cmsketch, the upper-bound is probabilistic with $\delta$.
Thus, decrementing the count by any amount, for both \worstcase and \averagecase, can cause a loss of information.
Moreover, because the technique of Graphene cannot be adopted for CountMin-Sketch, it may lead to continuously triggering mitigation once any count gets large.
This is overcome by architectural optimizations by CoMeT~\cite{hpca-2024-comet}, which we discuss in \S\ref{subsec:4_architectural_optimization}.

\noindent
\textbf{ii) Throttle.}
Throttle as a mitigative action~\cite{hpca-2021-blockhammer} does not affect \freal of the mitigated row.
Thus, there is no need to decrease the count or \fest value by any amount.
If we desire deterministic security, algorithms such as \cmsketch, \lossycounting, and \misragries can be utilized, which have a deterministic \fest lower-bound.

\noindent
\textbf{iii) Shuffle.}
Shuffling does not affect \freal, either.
Thus, theoretically, algorithms such as \misragries, \lossycounting, and \cmsketch can be utilized.
However, \cmsketch does not directly support identifying the address of the heavy-hitter,\footnote{An additional structure such as Max-Heap can be added to support such operations~\cite{Algorithms-2005-countmin-sketch}.} undesirable for shuffling judiciously.
Similar to the technique used in Graphene, one can only trigger shuffling at every integer multiple of a certain threshold, which is something that \misragries/\spacesaving and \lossycounting can support when correctly configured.

%-------------------------------------------------------------------------------
\subsection{Architecture Support for Streaming Algorithm}
\label{subsec:4_architectural_optimization}
%-------------------------------------------------------------------------------

While the security of RH solutions relies on a tight \worstcase bound, architectural techniques can improve the performance in the \averagecase.

\noindent
\textbf{1) Cache.}
Traditionally, an architectural cache serves a similar, yet different purpose compared to a heavy-hitter streaming algorithms.
Heavy-hitter streaming algorithms aim to find the items with the highest frequencies from the \emph{past}.
By contrast, the goal of the cache is to preserve items that will be frequently used in the \emph{future}, as the oracle Belady's MIN~\cite{sj-1966-belays-min} suggests.

Cache structures can be synergistically used with streaming algorithms to mitigate the performance overhead of RH solutions~\cite{cal-2015-CRA,isca-2022-hydra,hpca-2024-comet}.
While such performance mitigation can be hindered when an attacker tries to cause a performance attack, their effectiveness is often preserved in the \averagecase with benign workloads.
For example, Hydra~\cite{isca-2022-hydra} uses a Group-Counter Table (GCT), which can be considered as a single hash CountMin-Sketch in retrospect, augmented by a cache structure and per-row DRAM counter.
Hydra is unique in that both CountMin-Sketch (GCT) and the cache serve as a filter to mitigate performance overhead, while the per-row DRAM counter serves the role of guaranteeing security. 
CoMet~\cite{hpca-2024-comet} formally adopts CountMin-Sketch for security, and introduces a sophisticated Recent Aggressor Table (RAT), which acts as a cache that prevents mitigative refresh from being triggered too often.
Both works incur small performance overheads in various scenarios.
Similar techniques can be used in the future for other RH solutions with different streaming algorithms or mitigation methods.

\noindent
\textbf{2) Anomaly Detection.}
The security of the RH defenses that solely relies on anomaly detection~\cite{arxiv-2025-marc,date-2022-learning} is ambiguous.
Also, while it is possible to pursue full security through ML-based anomaly detection~\cite{micro-2022-evax}, it is unclear if it will be more efficient compared to streaming algorithm-based defenses.
At the same time, leveraging anomaly detection to loosely identify attackers and take action to improve \averagecase performance can be promising~\cite{micro-2024-breakhammer}.

\noindent
\textbf{3) Spatial Variation.}
So far, we implicitly assumed with an abstraction that all rows are equally vulnerable to RH, dictated by a certain \rhth number.
However, this is not true~\cite{hpca-2024-spatial, micro-2021-uncovering, micro-2022-hira, cal-2023-x-ray, isca-2020-revisiting}.
Sv\"ard~\cite{hpca-2024-spatial} demonstrates and exploits this fact to improve the existing RH solutions.
Although not an architectural defense solution, TAROT~\cite{asplos-2024-tarot} also exploits this observation to implement a low-cost software-based RH solution.
Exploiting the spatial variation, future streaming algorithms and solutions can adjust the weight of each activation based on the vulnerability of its victim rows to RH bitflips.

\noindent
\textbf{4) Sliding Window.}
RH defense solutions are conservative when considering auto-refresh.
For example, considering that the refresh cycle of a certain row can be pathologically asynchronous to the streaming algorithm counter table reset, we set the target \rhth to be halved~\cite{micro-2020-graphene}.
This results in a two-fold degradation in the area, energy, and performance overhead of various RH solutions.
While one study~\cite{hpca-2022-mithril} adopts an algorithmic optimization with a rolling counter to avoid table reset, such a technique is not universally applicable to other algorithms.
Sliding window algorithms~\cite{PODS-2004-approximate-counts-Arvind} hold the potential to improve such inefficiency, as gradually updating the counter table can prevent total counter updates.

\noindent
\textbf{5) Compute Express Link (CXL).}
A third-party (from the perspective of a processor) memory controller at a CXL memory device builds upon the sequential interface of PCIe to link a processor with memory~\cite{micro-2023-demystifying}.
This enables i) an additional layer of request scheduling, ii) logic process technology for an RH solution, and iii) non-deterministic timing in memory accesses.
When its memory request scheduling is conducted more aggressively to reduce the row-buffer hit rate, the stream length $N$ decreases.
This reduces the attack success probability (Figure~\ref{fig:reservoir-mint}), area/capacity requirements of RH solutions, and false-positive rates (Table~\ref{tbl:algorithms}).
Implementing RH solutions at the CXL controller allows RH solutions to leverage the logic process and utilize DRAM address mapping simultaneously.
Finally, non-deterministic timing significantly hampers attackers from acquiring address mapping, which is often considered to be easily compromised in the threat models of architectural RH defenses.

\begin{figure*}[!tb]
  \center
  \vspace{0.0in}
  \includegraphics[width=0.89\linewidth]{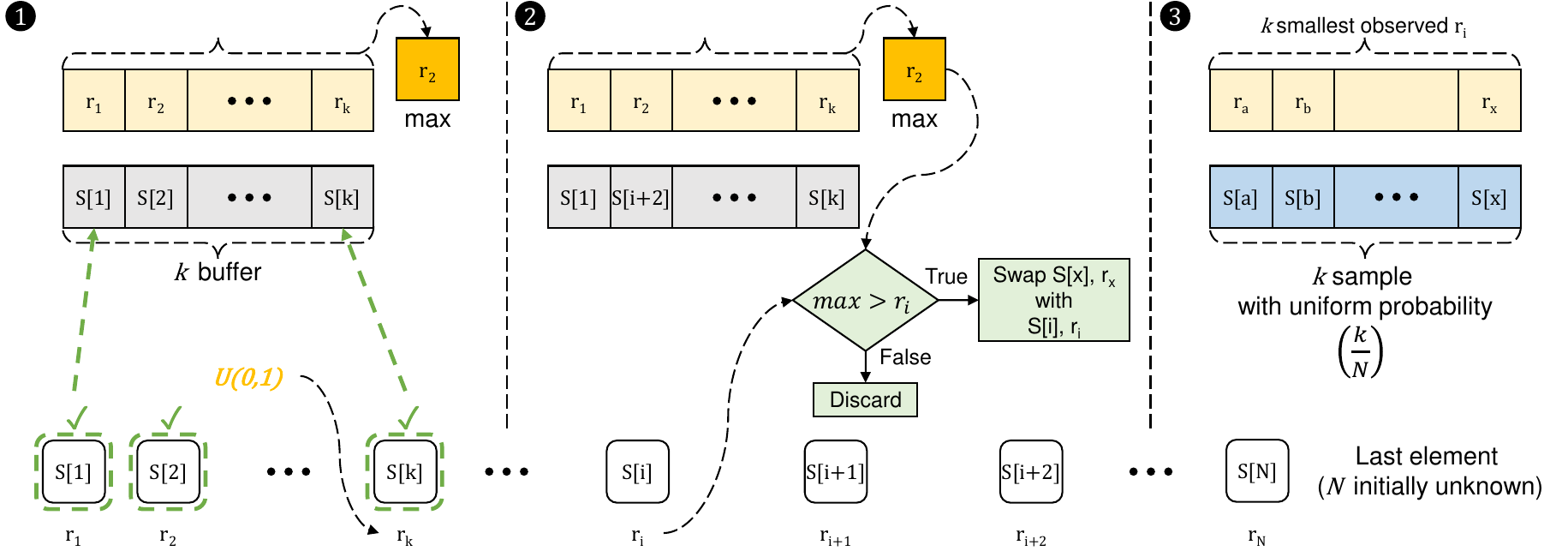}
  \vspace{-0.15in}
  \caption{Hardware friendly version~\cite{JASA-1962-Method-4} of \reservoir.
  }
  \vspace{-0.0in}
  \label{fig:method-4}
\end{figure*}

%-------------------------------------------------------------------------------
\section{Case Studies of Prior Defenses and Streaming Algorithms}
%-------------------------------------------------------------------------------
\label{sec:check_existing_algorithm}

We provide two case studies of recent RH defenses (\S\ref{subsec:pride_mint} and \S\ref{subsec:dsac_proteas}) and their associated streaming algorithms (\S\ref{subsec:reservoir_sampling} and \S\ref{subsec:sticky_sampling}) that were overlooked.
Prior to understanding the connection between streaming algorithms and architectural RH defenses~\cite{micro-2020-graphene}, defenses relied on novel yet less optimal methods or reinvented the existing algorithms.
Following the introduction of the link between two,
studies have begun adopting well-studied algorithms~\cite{hpca-2022-mithril,sp-2022-protrr,hpca-2021-blockhammer,hpca-2024-comet,isca-2022-hydra,asplos-2022-rrs}, achieving high efficiencies in various settings.
However, some RH defense studies still overlook\footnote{The fact that Misra-Gries~\cite{scp-1982-misragries} algorithm has been reinvented multiple times~\cite{TODS-2003-Misra-Gries-reinvent1, ESA-2002-misra-gries-reinvent2, ICDT-2005-Space-Saving} highlights the difficulty in avoiding such pitfalls.
} related algorithms, resulting in solutions that are overly conservative (\S\ref{subsec:pride_mint}) or even insecure (\S\ref{subsec:dsac_proteas}).

\begin{table}[!tb]
    \centering
    \small
    \caption{Symbols in \S\ref{sec:check_existing_algorithm}
    }
    \label{tbl:symbols_adversarial}
    \Scale[0.95]{
    \begin{tabular}{p{0.13\columnwidth} p{0.82\columnwidth}} 
        \Xhline{3\arrayrulewidth}
        \textbf{Symbol} & \textbf{Description} \\ 
        \Xhline{1.5\arrayrulewidth}
        $S[\ ]$   & Stream. $S[i]$ indicates $i^{th}$ element in a stream. \\
        $k$     & Target number of final samples from a stream. \\
        $N$     & Total length of a stream. \\
        \psample    & Sampling probability. \\
        \minentry   & Minimum number of entries in \spacesaving \\
        \Xhline{0.5\arrayrulewidth}
        \Xhline{3\arrayrulewidth}
    \end{tabular}
    }
    \vspace{-0.08in}
\end{table}

%-------------------------------------------------------------------------------
\subsection{Case-1 Algorithm: \reservoir}
\label{subsec:reservoir_sampling}
%-------------------------------------------------------------------------------

Recent proposals for sampling-based RH protection schemes implemented at the DRAM-level~\cite{hpca-2022-mithril,isca-2024-pride,MICRO-2024-mint} overlook the well-established \reservoir algorithms~\cite{TOMS-1985-algorithm-Z}. 
\reservoir~\cite{TOMS-1985-algorithm-Z} \emph{uniformly} samples a fixed number of items from a stream of \emph{unknown length}.
The processor-side sampling method can be straightforward because it can trigger mitigation whenever a sampling occurs, based on Bernoulli sampling~\cite{isca-2014-flipping}.
By contrast, a DRAM-side mechanism must uniformly sample a fixed number of items (\eg, one) from a stream of unknown length within each \texttt{REF} interval (\ie, every tREFI 3.9\us).
For example, while there can be a maximum of 73 ACTs between two \texttt{REF}s (when tRC is 48ns and tREFI is 3.9\us.), the exact number of ACTs that will come in is unknown in advance.
This is not the case for RFM, where the \texttt{RFM} itself is generated at every fixed number of ACTs.

\noindent
\textbf{Operation.}
Among various versions of \reservoir~\cite{TOMS-1985-algorithm-Z, TOMS-1994-algorithm-L}, Figure~\ref{fig:method-4} shows one optimization that is hardware-friendly.
\blackcircled{1}
First, initial $k$ items are inserted into the $k$-sized buffer.
At the same time, $k$ random numbers in range $U[0,1]$ are generated and stored in a separate array.
Also, we find the maximum among the random numbers and store its value and index.
\blackcircled{2}
For each newly encountered item, we generate a new random number, again in the range $U[0,1]$.
Then, we compare the newly generated random number with the current maximum random number.
If the newly generated random number is smaller than the current maximum, we replace both the random number and the associated sampled element (\eg, $r_2$ and $S[2]$).
\blackcircled{3}
At the end of the stream, $k$ samples are uniformly acquired.
The final $k$ random numbers are $k$ smallest random numbers among all the observed ones.

\begin{figure*}[!tb]
  \center
  \vspace{0.0in}
  \includegraphics[width=0.86\linewidth]{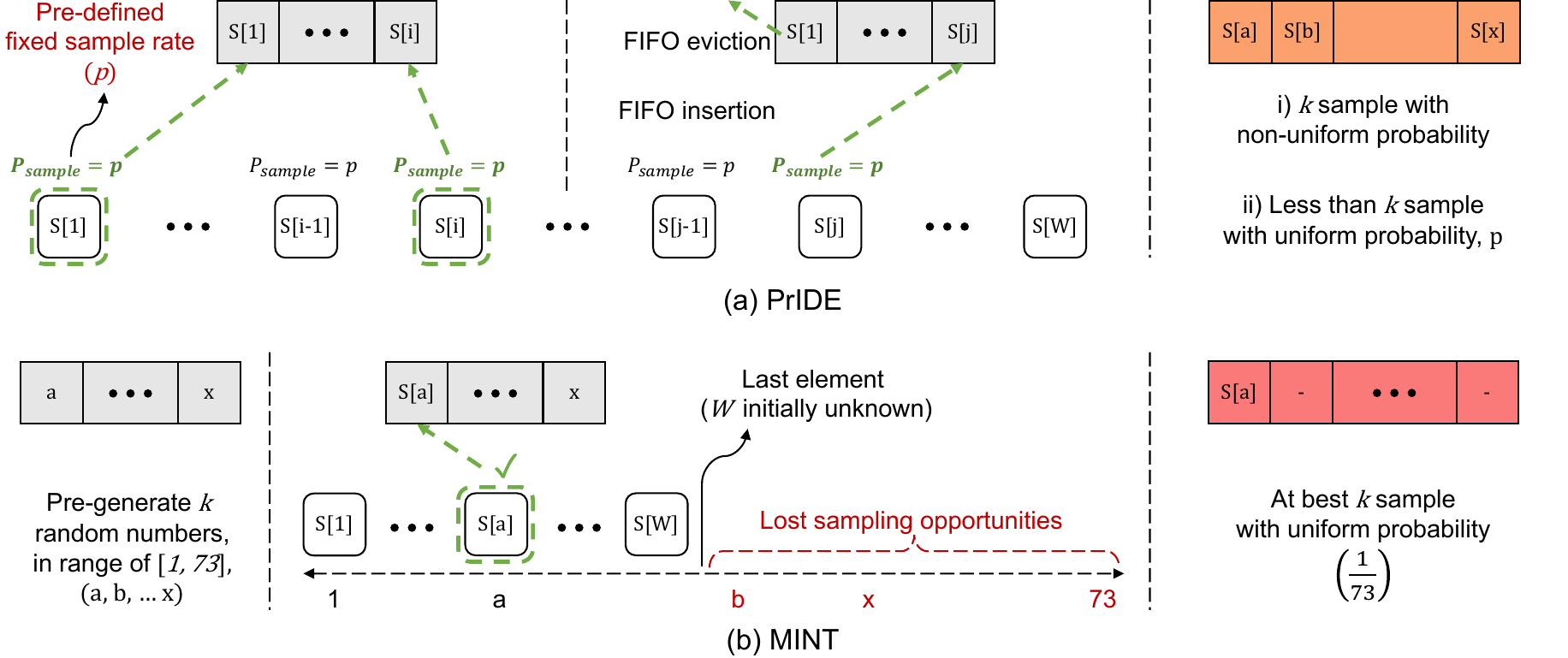}
  \vspace{-0.06in}
  \caption{Overview of (a) PrIDE~\cite{isca-2024-pride} and (b) MINT~\cite{MICRO-2024-mint}. 
  }
  \vspace{-0.06in}
  \label{fig:pride}
\end{figure*}

\begin{figure}[!tb]
  \center
  %\vspace{0.0in}
  \includegraphics[width=0.85\linewidth]{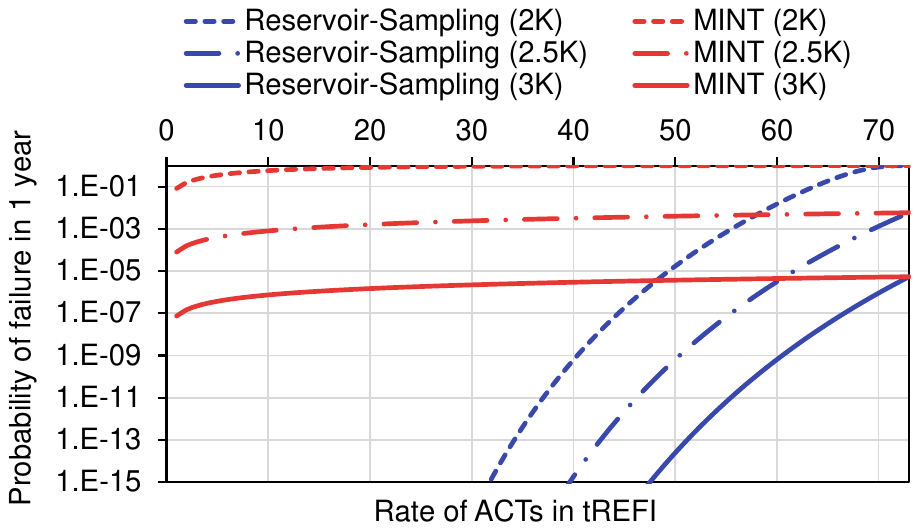}
  \vspace{-0.05in}
  \caption{Probability of failure of \reservoir and MINT across various number of ACTs in tREFI and \rhth of 2K, 2.5K, and 3K. The lower, the more secure.}
  \vspace{-0.05in}
  \label{fig:reservoir-mint}
\end{figure}

%-------------------------------------------------------------------------------
\subsection{Case-1 Defenses: Sampling-based Defenses}
\label{subsec:pride_mint}
%-------------------------------------------------------------------------------

Multiple architectural studies have tackled the same problem of \reservoir when building a DRAM-side sampling-based solution.
However, they~\cite{hpca-2022-mithril,isca-2024-pride,MICRO-2024-mint} have overlooked \reservoir, either proposing less efficient algorithms or nearly rediscovering it.

\noindent
\textbf{PrIDE.}
PrIDE~\cite{isca-2024-pride} employs a structured approach by combining Bernoulli sampling with a FIFO buffer (Figure~\ref{fig:pride}(a)). 
Initially, each \texttt{ACT} (item) undergoes Bernoulli sampling with a predefined fixed sampling rate ($P_{sample}=p$).
When an \texttt{ACT} is sampled, its row address is stored in a fixed-size buffer.
Each address in the buffer is then mitigated sequentially at every \texttt{REF}, following a FIFO order.
If the buffer overflows, addresses are evicted in a FIFO manner, with new addresses inserted at the tail.

However, because the number of \texttt{ACT}s and the ACT rate are unpredictable, a fixed sampling rate may be inappropriate.
This forces PrIDE to adopt an overly conservative approach, reducing its efficiency.
When the sampling rate is too high relative to the access stream, buffer overflows can occur frequently.
Conversely, when the rate is too low, the buffer may often be empty, missing mitigation opportunities during refresh periods.
Further, as noted in more recent work~\cite{MICRO-2024-mint}, each \texttt{ACT} within a tREFI window does not have a uniform probability of being selected for refresh.

\noindent
\textbf{MINT.}
MINT~\cite{MICRO-2024-mint} takes an approach by choosing random numbers in advance from 1 to the maximum number of ACTs within a tREFI window (\eg, 73).
This method was already implied in PARFM~\cite{hpca-2022-mithril}; however, MINT formalized the security analysis for the first time.
First, MINT generates $k$ random numbers, where $k$ is the number of mitigations that can occur at every \texttt{REF}.
Then, for each random number $rand$, the $rand^{th}$ \texttt{ACT} (\ie, $S[rand]$) is selected and its address is buffered.
All the buffered addresses are mitigated at the following \texttt{REF} (Figure~\ref{fig:pride}(b)).

This method only guarantees a uniform sampling of $k$ samples when the ACT stream occurs at the \emph{maximum rate} (\eg, 73 ACTs per tREFI).
When the stream length of ACTs within tREFI is shorter, fewer elements are sampled.
This leads to a loss of mitigation opportunity on the subsequent \texttt{REF}.
When the origin of the ACT stream is solely the attacking adversary, we could only assume the maximum rate case.
As MINT demonstrates, it is most efficient when the attacker generates ACTs at the maximum frequency.
The same applies even in the hypothetical case where \reservoir is applied instead of MINT, as a lower rate of ACT leads to a higher chance of being mitigated.

MINT can be less efficient in the \averagecase where benign workloads are accessing DRAM.
Figure~\ref{fig:reservoir-mint} demonstrates how the probability of failure of MINT and that of \reservoir differ as the length of the ACT stream in tREFI changes.
Even though the security guarantee in the \worstcase is identical between MINT and \reservoir, RH bitflips and ensuing silent data corruption can be further suppressed by \reservoir.
Moreover, if the attacker fails to generate the maximum rate of ACTs due to various causes (\eg, inter-process interference~\cite{isca-2008-parbs}, page policy~\cite{intel-2024-pagepolicy,micro-2011-minimalist}, or memory controller scheduling~\cite{isca-2000-fr-fcfs,isca-2008-parbs,micro-2007-stfm,micro-2006-fair-queuing,micro-2010-TCM}), the attack success probability quickly degenerates.

It is also unclear if MINT is hardware efficient.
MINT requires generating uniform random numbers in an odd range (\eg, 73), which also varies when the temperature or the number of DIMMs per channel changes.
\reservoir in Figure~\ref{fig:method-4} only requires random number in power-of-two ranges.

%-------------------------------------------------------------------------------
\subsection{Case-2 Defenses: Sampling augmented Tracker -based Defenses}
\label{subsec:dsac_proteas}
%-------------------------------------------------------------------------------

For case two, we discuss the RH defenses first.
If possible, adopting sampling-based randomness into counter-based streaming algorithms and the RH solution can provide a knob to balance between performance and area overhead.
Efforts to adopt sampling-based randomness into tracker-based RH defenses were proposed early on (\eg, MRLoc~\cite{dac-2019-mrloc} and ProHIT~\cite{dac-2017-prohit}).
Such prior studies were perceived at the time to be secure and efficient against benign and synthetic access patterns.
However, unlike pure random sampling~\cite{isca-2014-flipping}, such sampling-augmented counter-based solutions were later proven to be insecure against tailored adversarial patterns~\cite{micro-2020-graphene}.

\noindent
\textbf{DSAC.}
In line with such prior art, DSAC~\cite{arxiv-2023-dsac} proposed a solution that builds upon the \spacesaving algorithm with sampling-based randomness.
We refer the reader who is unfamiliar with the \spacesaving (\misragries) algorithm and its adversarial patterns to Appendix~\ref{sec:appendix}.
DSAC differs with \spacesaving in two aspects.
First, DSAC \emph{invalidates} a counter table entry at every mitigation, once per \texttt{REF}; this is based on the authors' description that an entry ``\emph{can be reset to 1 after TRR}''~\cite{arxiv-2023-dsac}.
Second, DSAC does not blindly swap the new address with the \minentry entry when an element misses the counter table.
Instead, DSAC \emph{stochastically} swaps the new address and increments the value.
The probability of swap is inversely proportional to the current value of \minentry (\ie, $1/(\texttt{Min}+1)$).

DSAC claimed that this optimization provides robustness against the ``thrashing'' adversarial patterns of \spacesaving.
Through such a technique, DSAC aims to provide a probabilistic, yet strong guarantee while reducing the area overhead of \spacesaving.
However, although the reason was unclear, it was proven vulnerable when tested  with fuzzers~\cite{isca-2024-pride}, which we also concurred.
We further discuss their adversarial activation patterns for each of their technique, for the first time.

Henceforth, we denote the aggressor rows as $r_{i}$, where $i$ is the row address of the aggressor.
Each pattern is denoted with parentheses and superscript; for example, $(r_1,r_2)^2$ indicates an activation pattern of $r_1 \rightarrow r_2 \rightarrow r_1 \rightarrow r_2$.

\begin{figure}[!tb]
  \center
  \vspace{0.0in}
  \includegraphics[width=0.98\linewidth]{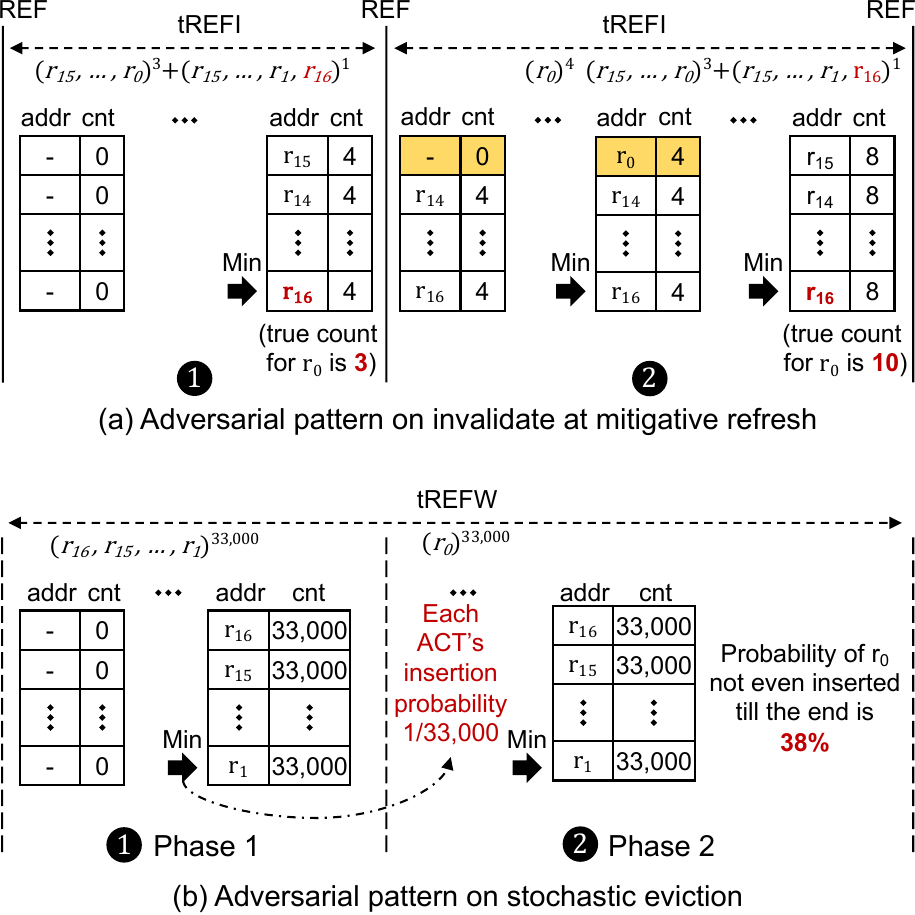}
  \vspace{-0.06in}
  \caption{An exemplar case of adversarial patterns of (a) invalidation at a mitigative refresh and (b) stochastic eviction.
  }
  \vspace{-0.06in}
  \label{fig:dsac}
\end{figure}

\noindent
\textbf{DSAC adversarial - i) post-mitigation invalidation.}
Figure~\ref{fig:dsac}(a) illustrates an adversarial pattern when \spacesaving is modified to \emph{invalidate} an entry that is mitigated at each \texttt{REF}.
In the following example, we consider a 16-entry counter table.
\blackcircled{1}
First, identical to the adversarial pattern of \spacesaving, 16 different rows are activated in a round-robin manner (\eg, $(r_{15}, r_{14},\ldots,r_0)^3$).
Right before an \texttt{REF}, 15 different rows and 1 new row are activated (\eg, $(r_{15}, r_{14},\ldots,r_1,r_{16})^1$).
Then, at the \texttt{REF}, one entry will be mitigated.
In DSAC with post-mitigation invalidation, the count value of such an entry will become 0.
\blackcircled{2}
During the second tREFI, however, our target row $r_0$ can be initially activated in a burst (\eg, $(r_0)^4$),
which can be as long as the second minimum count (\eg, 4).
Only after this burst, a total of 16 row addresses can be again equally activated (\eg, $(r_{15}, r_{14},\ldots,r_0)^4$).
Such a burst on $r_0$ can intensify as the pattern is repeated over time as the second minimum value gradually increases.
This results in a \emph{worse} bound compared to the \spacesaving algorithm.

\noindent
\textbf{DSAC adversarial - ii) stochastic eviction.}
For stochastic eviction (without post-mitigation invalidation), the attacker's goal is to skew the counter-table's state. 
If successful, a specific row (\eg, $r_0$) can be penalized with a low swapping probability.

The attacker can execute the following phase-changing attack (Figure~\ref{fig:dsac}(b)).
\blackcircled{1}
During phase 1, 16 different row addresses are activated (\eg, $(r_{16},r_{15},\ldots,r_{1} )$), each with a large number of activations (\eg, 33,000).
Although multiple \texttt{REF}s and the ensuing mitigations occur during this period, the \minentry value is forced to increase (\eg, 33,000), as the count of the mitigated entry cannot be decremented below \minentry.
\blackcircled{2}
Then, during phase 2, a new row address (\eg, $r_0$) is activated, solely monopolizing all the activations for the rest of the tREFW interval (\eg, 33,000).
During the second phase, there is only a small chance that the new address ($r_0$) is inserted into the table because the \minentry value is already too large.
In our example, the probability of $r_0$ never being inserted into the counter table until the end of tREFW is 0.38.
When the attack persists for one minute, the probability that $r_0$ reaches 33,000 ACT count without mitigation is 1.0.

When the invalidation techniques on mitigation and stochastic eviction are combined, the security is rather ambiguous.
Just as ProHIT~\cite{dac-2017-prohit} and MRLoc~\cite{dac-2019-mrloc} were proven vulnerable, results with the fuzzer~\cite{isca-2024-pride} suggest that DSAC with both techniques combined still has a vulnerability.
While another recent study has proposed a sampling-augmented counter-based RH solution~\cite{arXiv-2024-proteas}, its security is still demonstrated only with results against certain adversarial patterns without rigorous proof.
The fact that its security converges to that of pure sampling~\cite{isca-2014-flipping,arXiv-2024-proteas} suggests not only its robustness but also indifference from prior sampling-based approaches~\cite{isca-2024-pride,isca-2014-flipping}.

Similarly, various sampling-based streaming algorithms~\cite{SIGCOMM-2002-Sticky-sampling-weaker,Infocom-2006-Sketch-guided-sampling} have been proposed in a different context such as networking, yet their tight \worstcase bounds are mostly unexplored.
This is reasonable as their primary target is to provide good bounds in \averagecase.
We posit that the vulnerabilities in ProHIT, MRLoc, and DSAC arose because these solutions did not rely on mathematically proven \worstcase bounds in the frequency estimation.
In the following \S\ref{subsec:sticky_sampling}, we discuss how to provide a tight bound when randomness is added to the counter-based algorithms while introducing the \stickysampling algorithm.
Yet, we do not come with promising news as the \worstcase area overhead of \stickysampling is inferior to other frequency estimation algorithms.

%-------------------------------------------------------------------------------
\subsection{Case-2 Algorithm: \stickysampling}
%-------------------------------------------------------------------------------
\label{subsec:sticky_sampling}

As an example of a counter-based algorithm with a sampling technique, we introduce the \stickysampling algorithm~\cite{VLDB-2002-Lossy-sticky}.
The distinguishing parts of the algorithm are that i) the sampling rate is adjusted continuously, and ii) there exists a compression phase.
The algorithm provides a tight \worstcase bound, which prior related RH defenses~\cite{dac-2017-prohit,dac-2019-mrloc,arxiv-2023-dsac,arXiv-2024-proteas} did not provide.

\noindent
\textbf{\stickysampling.}
\stickysampling consists of two subroutines, \update and \compress , which are repeated across the whole stream.
We provide the pseudo code of \update/\compress (Algorithm~\ref{algo:ss-subroutines}) and the overall algorithm (Algorithm~\ref{algo:sticky-sample}) in Appendix~\ref{sec:appendix}.
Prior to both subroutines, parameters $\epsilon$ and $\delta$ are set.
This determines the parameter $t$, which defines the window width; \stickysampling divides the incoming stream of elements into windows, and performs the \compress subroutine at every window boundary.

\noindent
\textbf{\update.}
When a new element hits an entry in the \stickysampling counter structure ($C\{\}$), the corresponding count is incremented by one.
If a miss, it is inserted to $C\{\}$ with a sampling rate \psample.
While this is similar to the prior studies~\cite{SIGCOMM-2002-Sticky-sampling-weaker, arXiv-2024-proteas}, \psample is continuously adjusted mathematically.

\noindent
\textbf{\compress.}
At every window boundary, \compress is triggered.
For each counter entry in $C\{\}$, an unbiased coin is flipped repeatedly until heads come out.
Each counter is decremented by the total number of observed tails until a head appears, based on each's own coin toss.
One can also simply utilize a random number generator for a geometric distribution instead~\cite{TOMS-1985-algorithm-Z}.
Following the counter decrement, an entry is invalidated if the count becomes zero.
After each \compress subroutine, the width of the window is doubled and \psample is halved.

The \compress subroutine prevents the number of entries in $C\{\}$ from exploding.
After the \compress subroutine, $C\{\}$ becomes \emph{the exact state it would have been, had it been sampling at the new rate from the beginning}~\cite{VLDB-2002-Lossy-sticky}.
This means that \stickysampling is robust against adversarial patterns, equivalent to pure Bernoulli- and \reservoir.

\section{Conclusion}
\label{sec:8_conclusion}
In this paper, we have systematized a decade of architectural RowHammer defense solutions.
We also analyzed the tracker-based and tracker-less defenses through the lens of streaming algorithms.
We provided a guide on which algorithm to utilize upon various circumstances, including target \rhth, location, and mitigative action.
Further, we offered a guide to check on existing algorithms when architecting RowHammer defenses.
We demonstrated how \reservoir can improve the related RowHammer defenses, and introduced \stickysampling.

\section*{Acknowledgements}
We thank the anonymous reviewers for their constructive comments and suggestions.
This research was in part supported by Institute of Information \& communications Technology Planning \& Evaluation (IITP) grant funded by the Korea government (MSIT) [RS-2021-II211343, RS-2023-00256081, RS-2024-00456287],
by the National Research Foundation of Korea (NRF) grant funded by MSIT (No. RS-2024-00405857), and a grant from PRISM, one of the seven centers in JUMP 2.0, a Semiconductor Research Corporation (SRC) program sponsored by DARPA.
This work was done when Michael Jaemin Kim was at Seoul National University (SNU).
Nam Sung Kim and Jung Ho Ahn have a financial interest in Samsung.
Jung Ho Ahn, the corresponding author, is with the Department of Intelligence and Information and the Interdisciplinary Program in Artificial Intelligence, SNU.


%\balance
%-------------------------------------------------------------------------------
\bibliographystyle{IEEEtranS}
\bibliography{ref}

%\clearpage
\appendices
\counterwithin{figure}{section}
\renewcommand{\thealgorithm}{A.\arabic{algorithm}}
\begingroup
  % --- float spacing just for the appendix ---
  \setlength{\textfloatsep}{4pt}
  \setlength{\floatsep}{4pt}
  \setlength{\intextsep}{4pt}
  \section{}
\label{sec:appendix}

%-------------------------------------------------------------------------------
\subsection{\misragries/\spacesaving}
\label{subsec:misra-gries}
%-------------------------------------------------------------------------------

\noindent
\textbf{\spacesaving.}
\spacesaving algorithm~\cite{ICDT-2005-Space-Saving} is a deterministic counter-based algorithm that is isomorphic to \misragries~\cite{scp-1982-misragries}.
\spacesaving has a counter table where each entry stores an address and a corresponding counter.
When an element in a stream hits the counter table, its count is incremented.
Whenever a new element misses the table, an existing entry having the \emph{minimum} count value (\minentry) is swapped. 
Figure~\ref{fig:misra-gries-dsac-proteas} illustrates the operation of \spacesaving.

\noindent
\textbf{\spacesaving adversarial.}
An adversarial pattern of \spacesaving is as follows (Figure~\ref{fig:misra-gries}).
Consider a case where there are 16 counters in the \spacesaving table.
In this case, 17 different row addresses are equally activated (\eg, $(r_{0},r_{1},\ldots,r_{16})^4$).
After the accesses, an address (\eg, $r_0$) is missed on the table, although its true activation count is 4, an equal value to \minentry.
Then a REF command is issued, mitigating one address from the table (in this example, only a single address is mitigated per REF).
The same pattern is repeated for the rest of the tREFW period.

Even after the mitigation, its counter value cannot be invalidated or zeroed.
Although the accumulated number of ACTs on the aggressor row (\eg, $r_0$) is zero after the mitigation, we do not know how much of the counter value in the \spacesaving count table is attributed to the mitigated address.
If we na\"ively decrease the count value, it can cause unwanted loss of information on counts that belong to the addresses that are off the \spacesaving count table.
A safe option is not to decrement~\cite{micro-2020-graphene}, or to decrement to \minentry~\cite{hpca-2022-mithril,sp-2022-protrr}.
%
%This is well considered in prior literatures~\cite{micro-2020-graphene, hpca-2022-mithril, sp-2022-protrr}.
%
Post-mitigation counter update is one of the unique problems that does not exist in the streaming algorithm context, to the best of our knowledge.

\begin{figure}[!tb]
  \center
  %\vspace{0.0in}
  \includegraphics[width=0.9\linewidth]{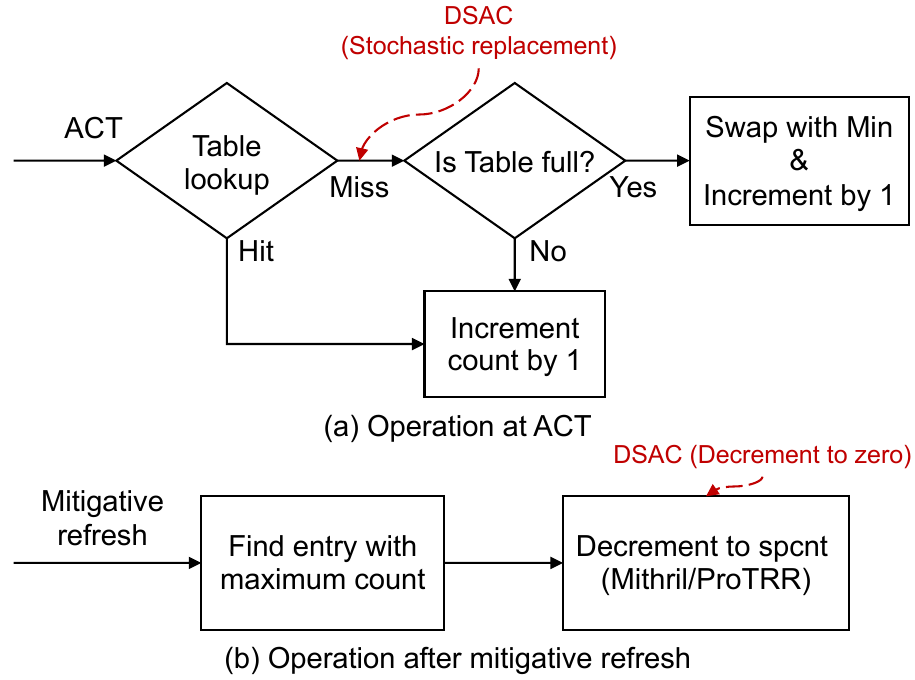}
  \vspace{-0.1in}
  \caption{\spacesaving at every ACT (a) and mitigative action (b), including the modifications by DSAC~\cite{arxiv-2023-dsac}.}
  \vspace{-0.0in}
  \label{fig:misra-gries-dsac-proteas}
\end{figure}

\begin{figure}[!tb]
  \center
  \vspace{0.0in}
  \includegraphics[width=0.98\linewidth]{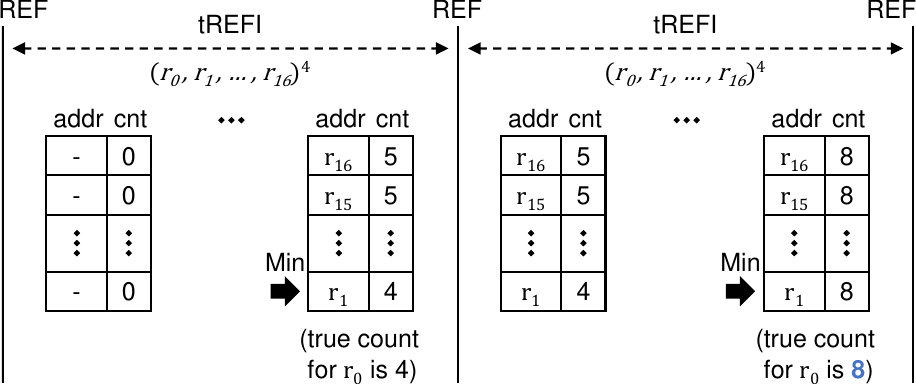}
  \vspace{-0.1in}
  \caption{Adversarial patterns of \spacesaving.}
  \vspace{-0.0in}
  \label{fig:misra-gries}
\end{figure}

%-------------------------------------------------------------------------------
\subsection{\stickysampling}
\label{subsec:sticky_sampling_appendix}
%-------------------------------------------------------------------------------

\begin{algorithm}[tb!]
\caption{\stickysampling~\cite{VLDB-2002-Lossy-sticky}}
\label{algo:sticky-sample}
\small
\begin{algorithmic}[1]
\State \textbf{Input:} $\epsilon$, $\delta$, $C\{\}$
\State \textbf{Output:} Approximate frequency counts of items in an input stream
\vspace{0.2cm}
\State Initialize:
\State \hspace{0.5cm} $processed \gets 0$ \hfill // Total number of items processed
\State \hspace{0.5cm} $t \gets \lceil \frac{1}{\epsilon} \ln \frac{1}{2\epsilon\delta} \rceil$ \hfill // $\epsilon$ is set as $\frac{supoprt}{2}$
\State \hspace{0.5cm} $window\_width \gets 2t$ \hfill
\State \hspace{0.5cm} $P_{sample} \gets 1$ \hfill
\State \hspace{0.5cm} $C \gets \emptyset$ \hfill // Counter map \{item $\to$ count\}
\vspace{0.2cm}
\For{each item $x$ in stream}
    \State $processed \gets processed + 1$ 
    \State \textbf{Update($x$)}
    \If{$processed = window\_width$}
        \State \textbf{Compress($C$)}
        \State $window\_width \gets 2 \times window\_width$
        \State $P_{sample} \gets P_{sample}/2$
    \EndIf
\EndFor
\State \textbf{Return:} $freq$
\end{algorithmic}
\end{algorithm}

\begin{algorithm}[tb!]
\caption{Subroutines of \stickysampling~\cite{VLDB-2002-Lossy-sticky}}
\label{algo:ss-subroutines}
\small
\begin{algorithmic}[1]
\Function{Update}{$x$}
    \If{$x \in C$}
        \State $C[x] \gets S[x] + 1$
    \Else
        \State Generate a random number $r \in [0, 1]$
        \If{$r \leq P_{\text{sample}}$}
            \State $C[x] \gets 1$
        \EndIf
    \EndIf
\EndFunction
\vspace{0.2cm}
\Function{Compress}{$C$}
    \For{each element $x \in C$}
        \State $tails \gets 0$ \hfill // Initialize tails counter
        \While{True}
            \State Flip a coin
            \If{coin lands on tails}
                \State $tails \gets tails + 1$
            \Else
                \State \textbf{break} \hfill // Stop flipping on first head
            \EndIf
        \EndWhile
        \State $C[x] \gets C[x] - tails$ \hfill // Decrement by tails
        \If{$C[x] \leq 0$}
            \State Remove $x$ from $C$
        \EndIf
    \EndFor
\EndFunction
\end{algorithmic}
\end{algorithm}

   % <— your appendix.tex
\endgroup

\clearpage
\newpage % The Meta-Review should at least start on a new column

% Use \appendices and not \appendix due to IEEEtran.cls quirks
% \appendices % if not used earlier

\section{Meta-Review}

The following meta-review was prepared by the program committee for the 2026
IEEE Symposium on Security and Privacy (S\&P) as part of the review process as
detailed in the call for papers.

\subsection{Summary}
The paper systematizes and taxonomizes 48 different architectural Rowhammer defenses. It outlines how these defenses are connected to many classical streaming algorithms. Additionally, it provides guidelines for architects of future Rowhammer defenses showing (a) which streaming algorithms to use for a defense given a certain set of wanted features, and (b) how insights from streaming algorithms can be used to improve existing defenses (or design new ones). In the end, the paper introduces two new Rowhammer defenses based on Reservoir-Sampling and Sticky-Sampling.

\subsection{Scientific Contributions}
\begin{itemize}
\item Addresses a Long-Known Issue
\item Provides a Valuable Step Forward in an Established Field
\end{itemize}

\subsection{Reasons for Acceptance}
\begin{enumerate}
\item The paper addresses a long-known issue. While the connection between Rowhammer defenses and streaming algorithm has been noted before, reviewers appreciated the retrospective highlighting streaming algorithms in prior work that predates the discovery of this connection. All reviewers found this connection very interesting as it could inform the design of more effective next-generation mitigations.
\item The paper provides a valuable step forward in an established field. The authors' systematization yields the concrete insight that reservoir-sampling is an overlooked algorithm that improves on previous approaches.
\end{enumerate}

%%%%%%%%%%%%%%%%%%%%%%%%%%%%%%%%%%%%%%%%%%%%%%%%%%%%%%%%%%%%%%%%%%%%%%%%%%%%%%%%
\end{document}